\documentclass[showpacs,pra,twocolumn,aps,superscriptaddress,floatfix]{revtex4-1}
 \usepackage{bm} \usepackage{amsmath} \usepackage{amssymb} \usepackage{latexsym}
 \usepackage{amsfonts} \usepackage{graphicx}
\renewcommand\Re{\operatorname{Re}}
\renewcommand\Im{\operatorname{Im}}

   \newcommand\romand{\mathrm{d}}

\newcommand{\bea}{\begin{eqnarray}}
      \newcommand{\ea}{\end{eqnarray}} \newcommand{\eea}{\end{eqnarray}}
       
\usepackage{color}
\newcommand{\ha}{\ensuremath{\hat{a}}}
\newcommand{\hpsi}{\ensuremath{\hat{\Psi}}}

\newcommand{\eqnref}[1]{Eq.~(\ref{#1})}
\newcommand{\figref}[1]{Fig.~\ref{#1}}
\makeatletter
\newcommand{\rmnum}[1]{\romannumeral #1}
\newcommand{\Rmnum}[1]{\expandafter\@slowromancap\romannumeral #1@}
\makeatother

\begin{document}
\title{Band structure loops and multistability in cavity-QED}

\author{B.\ Prasanna Venkatesh}
\affiliation{Department of Physics and Astronomy, McMaster University, 1280 Main
St.\ W., Hamilton, ON, L8S 4M1, Canada} 
\author{J.\ Larson}
\affiliation{Department of Physics, Stockholm University, Se-106 91, Stockholm, Sweden}
\author{D.\ H.\ J.\ O'Dell}
\affiliation{Department of Physics and Astronomy, McMaster University, 1280 Main
St.\ W., Hamilton, ON, L8S 4M1, Canada}

\begin{abstract}
We calculate the band structure of ultracold atoms located inside a laser-driven optical cavity. For parameters where the atom-cavity system
exhibits bistability, the atomic band structure develops loop structures akin to the ones predicted for Bose-Einstein condensates in ordinary
(non-cavity) optical lattices. However, in our case the nonlinearity derives from the cavity back-action rather than from direct interatomic
interactions. We find both bi- and tri-stable regimes associated with the lowest band, and show that the multistability we observe can be analyzed in terms
of swallowtail catastrophes. Dynamic and energetic stability of the mean-field solutions is also discussed, and we show that the bistable solutions
have, as expected, one unstable and two stable branches. The presence of loops in the atomic band structure has important implications for proposals concerning Bloch oscillations of atoms inside optical cavities [Peden \textit{et al}.,
Phys. Rev. A \textbf{80}, 043803 (2009), Prasanna Venkatesh \textit{et al}., Phys. Rev. A \textbf{80}, 063834 (2009)].                        
\end{abstract}
\maketitle

\section{Introduction}\label{sec1}

Optical bistability is a manifestation of nonlinearity in optical systems which is well known in the laser physics  community
\cite{gibbs76,bonifacio} (see also \cite{meystre} and \cite{gardiner} and references therein). It describes a situation in which there are  two
possible stable output light intensities for a single input intensity,  and occurs when an optical medium with a nonlinear refractive index is
placed inside an optical cavity formed from two mirrors. The bistable behaviour results from the combination of the nonlinearity of the medium
with the action of the feedback provided by the mirrors. 

A new addition to the family of systems displaying optical bistability has recently been demonstrated in experiments performed by the ultracold
atom groups at Berkeley \cite{gupta07} and the ETH \cite{brennecke08} who found optical bistability in systems comprising of vapors of ultracold
atoms trapped inside  optical cavities which are driven by laser light.  The atomic vapor acts as a dielectric medium and, despite being tenuous,
can significantly perturb the light in a cavity with a small mode volume and high finesse if the cooperativity $\mathcal{C}_{N}$ is in the regime
$\mathcal{C}_{N}\equiv N g_{0}^{2}/2\kappa \gamma>1$, where $N$ is the number of atoms, $g_{0}$ is the single photon Rabi frequency, $2 \gamma$
is the atomic spontaneous emission rate in free space, and $2 \kappa$ is the cavity energy damping rate. The perturbation of the light by the
atoms is nonlinear and distorts the cavity lineshape away from being a lorentzian which is symmetric about the resonance frequency into one with
an asymmetric shape. For large enough cooperativity the lineshape becomes folded over (see Fig.\ \ref{fig:nonlorent} below), so that for a
certain range of frequencies there are three possible output light intensities (two stable, one unstable) from the cavity for a single input
intensity. The experiments \cite{gupta07} and \cite{brennecke08} exhibited this optical bistability as a hysteresis effect seen by chirping the
laser frequency through the cavity resonance from above and below the resonance: a sudden jump in the intensity of light transmitted through the
cavity was observed which occurred at two different frequencies, depending upon the direction of the chirp. 

An important difference between traditional laser systems and the ultracold atom experiments \cite{gupta07,brennecke08} is the origin of the
nonlinearity. In the former case the nonlinearity of the medium occurs in its polarization response, i.e. it arises from the internal degrees of
freedom of the atoms. By contrast, in the ultracold atom experiments the detuning of the cavity from atomic resonance was large enough that the
polarization response was in the linear regime. The nonlinearity was instead due to the response of the centre-of-mass wave function of the
atoms: the atoms re-arrange their position distribution according to the balance between the dipole force applied by the intracavity light field
(which forms a periodic lattice) and their zero-point energy. As a consequence, the depth of the optical lattice that forms inside the cavity in
experiments like  \cite{gupta07} and \cite{brennecke08} is not fixed purely by the drive laser intensity, as is the case in standard optical
lattices made by interfering laser beams in free space. Rather, when $\mathcal{C}_{N}>1$ the depth of the lattice is sensitive to the spatial
distribution of the atoms trapped in the cavity, and, in turn, the atoms' center-of-mass wave function is sensitive to the lattice depth. This
feedback nonlinearity, which leads to different amounts of transmitted/reflected light for a given input intensity depending on the spatial
distribution of the atoms,  has  been previously employed to detect the presence of a single atom in a cavity \cite{hood98}, as well as to
monitor the motion of atoms trapped in cavities \cite{ye99}.  More recently, there has also been considerable theoretical interest in the effect
of the feedback nonlinearity upon the many-body quantum state of ultracold atoms in cavities \cite{maschler05,larson08,zhang09,nagy09,zhou09}. We
note, in particular, the theoretical work on self-organisation and related phenomena \cite{nagy08}, culminating in the experimental observation
of the Dicke quantum phase transition \cite{baumann10}.

Striking nonlinear phenomena also occur when ultracold atoms are trapped in standard ``fixed'' free-space lattices \cite{morsch06}. Of special
interest to us here are the curious swallowtail loops that occur in the band structure (energy versus quasi-momentum) of atomic Bose-Einstein
condensates (BECs) in one-dimensional optical lattices. These loops have been studied theoretically by a number of groups
\cite{wu02,machholm03,mueller02,smerzi02,pethick} in order to explain the breakdown in superfluidity observed in experiments where a BEC flows
through a lattice \cite{burger01,cataliotti03}. The loops correspond to multiple solutions for the atomic wave function within a single band for
a certain range of quasi-momenta.  They can occur either around the boundaries of the Brillouin zone or the center, depending upon the band and
the sign of the interactions. They manifest themselves physically via a dynamical instability that destroys the superflow. However, the
nonlinearity responsible for the swallowtail loops in the free-space lattices is provided by the interatomic interactions, which become important
at the densities required for Bose-Einstein condensation. The loops occur when the strength of the interactions is above a critical value
\cite{wu02,machholm03}, and therefore non-interacting atoms in an optical lattice do not display these instabilities. Our purpose in this paper is to
investigate whether the cavity feedback nonlinearity associated with optical bistability in ultracold atoms can also lead to loops in the atomic
band structure.  As we shall see, the answer to this question is in general affirmative, and so band structure loops appear to be a robust
phenomenon which appear whatever the source nonlinearity, although the structure and location of the loops does depend on the details of the nonlinearity.

One consequence of loops in the atomic band structure is a hysteresis effect \cite{mueller02} if the quasi-momentum is swept through the band,
and a consequent loss of adiabaticity even if the quasi-momentum is swept infinitely slowly \cite{wu00}. Recent experiments on a  two-site lattice have confirmed this scenario \cite{IBloch}.  These effects have implications for experiments
that use Bloch oscillations of cold atoms in optical lattices  for high precision measurements, for example to determine the fine structure
constant $\alpha$ \cite{battesti04}, to measure gravity \cite{roati04,ferrari06}, or to investigate Casimir-Polder forces \cite{carusotto05}.  The band
structure hysteresis is reminiscent of the hysteresis described above in the context of the optical bistability experiments \cite{gupta07} and
\cite{brennecke08}. Indeed, as we shall show in this paper, for atoms in an optical cavity the two effects are different sides of the same coin, one being seen in the light and the other in the atoms. Of particular relevance, therefore, are two recent proposals
\cite{peden09,bpv09} concerning Bloch oscillations of atoms inside optical cavities that rely upon the modification of the transmitted/reflected
light arising from the feedback nonlinearity as a non-destructive measurement of the Bloch oscillations. The presence of loops will severely
disrupt the Bloch oscillations. In the case where the nonlinearity has its origin in atom-atom interactions the loops can be eliminated by, e.g.,
using spin polarized fermions \cite{carusotto05} or by reducing the interactions via a Feshbach resonance \cite{gustavsson08}. However, when the
nonlinearity is due to the feedback nonlinearity these methods do not apply, and one of our aims here is see if there are regimes where the
feedback nonlinearity is present and leads to a modification of the light, but remains below a critical value needed to form loops in the atomic
band structure.
In order to allow the origin of the nonlinearity to be clearly identified, we shall only consider non-interacting atoms in this paper, but our
calculations can be easily extended to include interactions.

The plan for the rest of the paper is as follows. In Section \ref{sec2} we give a brief description of the system and introduce the mean-field
hamiltonian describing cold atoms dispersively interacting with the single mode of a cavity. In Section \ref{sec3} we derive a reduced
hamiltonian describing the nonlinear evolution for the atomic field after the adiabatic elimination of the light field. In Section \ref{sec4} we
calculate the band structure by two different methods that solve for the spatially extended eigenstates (Bloch states) of the reduced
hamiltonian, and show that the two methods give the same results. We find that for certain parameter regimes, the energy as a function of
quasi-momentum develops loop structures. In
Section \ref{sec5} we recall the optical bistability in atom-cavity systems discussed by \cite{gupta07,brennecke08} and make the connection
between the loop dispersions and bistability. In Section \ref{sec6} we develop an analytical estimate for the critical pump strength necessary to generate loops and in Section \ref{sec7} we test this result by illustrating how the band structure depends on laser detuning and laser intensity, i.e.\ the birth and death of loops as these parameters are varied. In Section \ref{sec8}, we show that for quasi-momentum $q\neq 0$ the system can exhibit tristability (five solutions for
the steady state cavity photon number). 

A convenient mathematical framework for describing bifurcations of a system whereby there is a sudden change in the number of solutions as a parameter is smoothly changed is catastrophe theory \cite{thom,arnold,berry81,poston,nye99}. Catastrophe theory was applied to the problem of optical bistability in the late 1970s by a number of
authors including Gilmore and Narducci \cite{gilmore78} and Agarwal and Carmichael \cite{agarwal79}. In Section \ref{sec9} we relate the problem
of atom-cavity multistability to catastrophe theory and show that the system under consideration can be described by swallowtail catastrophes organized by an (unobserved) butterfly catastrophe and use this to understand multistability. This is followed in Section \ref{sec10} by a discussion of the stability of these solutions and finally
we conclude in
Section \ref{sec11}. There is also an appendix which summarizes some concepts of catastrophe theory.

\section{The hamiltonian}
\label{sec2}
Consider a gas of $N$ ultracold atoms inside a Fabry-Perot optical cavity. The atoms interact quasi-resonantly with a single cavity mode of
the electromagnetic field of frequency $\omega_{c}$, and it varies along the cavity axis as $\cos(k_{c}z)$, where $k_{c}=\omega_{c}/ c$.  This
cavity mode is externally pumped by a laser of frequency $\omega_{p}$. The relevant frequency relations can be characterized with the two
detunings
\begin{eqnarray}
\Delta_{c} & \equiv & \omega_{p}-\omega_{c}, \\
\Delta_{a} & \equiv & \omega_{p}-\omega_{a},
\end{eqnarray}
where $\omega_{a}$ is the atomic transition frequency. In the dispersive regime, the occupation of the excited atomic state is vanishingly
small and it can be adiabatically eliminated. A one-dimensional hamiltonian for the atom-cavity system in the dispersive regime can then be
written as \cite{maschler05,larson08}
\begin{align}
H &= -\hbar \Delta_{c} \ha^{\dagger} \ha + i \hbar \eta \left (\ha^{\dagger} -
\ha\right) \nonumber \\
 & + \int \romand z \ \hpsi^{\dagger} \left[-\frac{\hbar^2}{2M} \frac{\partial^2}{\partial z^2} +
\frac{\hbar g_{0}^2}{\Delta_a} \  \ha^{\dagger} \ha \cos^2 (k_{\mathrm{c}}z) \right] \hpsi,
\label{eq:hamiltonian}
\end{align}
where $\hpsi(z,t)$ and $\ha(t)$ are the field operators for the atoms and the cavity photons, respectively. We have written the
hamiltonian in a frame rotating with the pump laser frequency $\omega_{\mathrm{p}}$, which leads to the appearance of the two detunings. The
first term
is just the free field evolution of the cavity mode. The second term represents the laser coherently pumping the cavity at rate $\eta$, and the
third term describes the atomic part of the hamiltonian. It consists of a kinetic energy part and a potential energy part. The potential energy
term can either be understood as the atom moving in a periodic potential with amplitude $(\hbar g_{0}^{2} / \Delta_{a}) \langle \ha^{\dagger}\ha \rangle$, or, if
combined with the first term in the hamiltonian, as a shift in the resonance frequency of the cavity due to the coupling
between the atom and the field (see also Eq.~(\ref{eq:lighteqn}) below). 

The Heisenberg equations of motion for the atom and photon field operators can be
obtained from the above hamiltonian. In this paper we are interested in properties at a mean-field level, where operators are
replaced by their expectation values and correlations between products of operators are neglected. In other words, the light is assumed to be in
a coherent state with amplitude $\alpha(t) = \langle \ha \rangle$, and the atoms are assumed to all share the same single-particle wave function
$\psi(x,t) = \langle \hpsi  \rangle / \sqrt{N}$. From the Heisenberg equations these amplitudes obey the following coupled equations of motion
\cite{cct} 
\begin{align}
i  \frac{\partial \psi}{\partial t} &= \left(-\frac{\partial^2}{\partial x^2} +  U_0 \, n_{\mathrm{ph}} \cos^2(x) \right) \psi,
\label{eq:atomeqn}\\
\frac{\romand \alpha}{\romand t} &= \left(i \Delta_{c} - i NU_0 \int \romand x \cos^2(x)
\vert \psi(x) \vert^2 -\kappa \right) \alpha + \eta, \label{eq:lighteqn}
\end{align}
where we have scaled energies by the recoil energy $E_{\mathrm{R}} = \hbar^2 k_{\mathrm{c}}^2 / 2M$, 
and time by $\hbar/E_R$. The depth of the periodic potential seen by the atoms is then given by $ U_0 \, n_{\mathrm{ph}}$, where $n_{\mathrm{ph}}
\equiv \vert \alpha \vert^2$  is the mean number of photons in the cavity, and $U_{0}$ is defined as
\begin{equation}
U_{0}\equiv \hbar g_{0}^{2} / (\Delta_{a} E_{\mathrm{R}}) \ .
\end{equation}
The damping rate $\kappa$ (in units of $E_{\mathrm{R}}$) of the amplitude of the light field in the cavity has been added into the equation of
motion in order to account for a markovian coupling between the cavity mode and a zero temperature bath. We have also introduced the
dimensionless length $x
\equiv k_{\mathrm{c}} z$.  In the above equations, any fluctuations induced by the reservoirs have been neglected. This is justified when
considering relatively large quantum numbers, for corrections see reference \cite{nagy09}.

In this paper we are interested in the band structure and this requires the quasi-momentum to be a  good quantum number. Physically, this implies that
we are studying delocalized atomic wave functions  which cover many lattice sites (this is closer to the situation realized in  the experiment 
\cite{brennecke08} than that realized in \cite{gupta07} where the atoms were localized to just a few sites). We shall therefore expand the wave
function $\psi(x,t)$ in Bloch waves, as will be detailed in subsequent sections. It thus makes sense to normalize $\psi(x,t)$ over one period of
the potential as $\int_0^{\pi} \vert \psi \vert^2 dx = \pi $, and also evaluate the integrals appearing in the above equations over one period.
In particular, the integral in Eq.~(\ref{eq:lighteqn}) which determines the coupling between the atoms and the light will be defined by
\begin{equation}
\langle \cos^2(x) \rangle  \equiv \frac{1}{\pi} \int_0^{\pi} \vert \psi(x) \vert ^2
\cos^2(x) \romand x . 
\label{eq:coupling}
\end{equation}

\section{The reduced hamiltonian}
\label{sec3}
In this Section we shall eliminate the optical degrees of freedom from the hamiltonian 
in order to obtain a description only in terms of the atoms. This results in a nonlinear Schr\"{o}dinger equation and an energy functional we
shall refer to as the reduced hamiltonian.

Setting $\romand \alpha / \romand t = 0$ in  \eqnref{eq:lighteqn} gives the steady state light amplitude in the cavity,
\begin{align} 
\alpha &= \frac{\eta}{\kappa} \frac{1}{1+ i\frac{\Delta_{c} - NU_0
\langle \cos^2(x)\rangle}{\kappa}} \label{eq:stateqbm} 
\end{align}
which leads to the following equation for the steady state photon number
\begin{align}
n_{\mathrm{ph}} &= \frac{\eta^2}{\kappa^2 + \left(\Delta_{c} -
NU_0\langle \cos^2(x) \rangle \right)^2} \ .  \label{eq:selfconsphnum}
\end{align}
In fact, this expression also holds to high accuracy in many time-dependent situations because $\kappa$ is typically far greater than any
frequency associated with the evolution of the external degrees of freedom of ultracold atoms. The light field is then slaved to the atoms and
``instantaneously'' adjusts itself to any change in $\psi(x,t)$.
The steady state solution can then be substituted back into \eqnref{eq:atomeqn} to give us a single, closed, nonlinear Schr\"{o}dinger equation
for the atomic wave function
\begin{align}
i \frac{\partial \psi}{\partial t} = \left(-\frac{\partial^2}{\partial x^2} + \frac{U_0 \, \eta^2
\cos^2(x)}{\kappa^2 + \left(\Delta_{c} -
NU_0 \langle \cos^2(x) \rangle \right)^2}\right)\psi \ .  \label{eq:nonlineareqn}
\end{align}

The stationary solution $\psi(x,t) = \psi(x) e^{-i\mu t}$ of this equation leads to an expression for the eigenvalue $\mu$ of Eq.\ (\ref{eq:nonlineareqn}),
\begin{align}
\mu[\psi,\psi^{\ast}] = \frac{1}{\pi} \int_0^{\pi} dx \left( \left \vert \frac{d\psi}{dx}
\right \vert^2 + U_0 n_{\mathrm{ph}} \cos^2(x) \vert \psi(x) \vert^2 \right ).
\label{eq:chemicalpotl}
\end{align}
If the Schr\"{o}dinger equation (\ref{eq:nonlineareqn}) were linear, then the eigenvalue $\mu$ would be the energy of the atoms in state
$\psi$. Furthermore, the functional (\ref{eq:chemicalpotl}) would serve as the energy functional from which this Schr\"{o}dinger  equation could
be obtained as an equation of motion using Hamilton's equation
\begin{equation}
i \hbar \frac{\partial \psi}{\partial t}= \frac{\delta E[\psi, \psi^{\ast}]}{\delta \psi^{\ast}} \ . \label{eq:hamilton}
\end{equation}
i.e.\ $E[\psi, \psi^{\ast}]=\mu[\psi,\psi^{\ast}]$ for a linear equation.
However, this is not the case here. Instead, the eigenvalue $\mu$ is the chemical potential which is related to the true energy $E$ via
$\mu = \partial E / \partial N$ (a prominent example that illustrates this distinction is the Gross-Pitaevskii equation and its energy functional
\cite{pethick}). Using this fact, one can show that the true energy functional from which the equation of motion (\ref{eq:nonlineareqn}) can be
derived is in fact \cite{larson08,zhou09}
\begin{align}
 E[\psi] &= \frac{N}{\pi} \int_0^{\pi} dx \left \vert \frac{d\psi}{dx}
\right \vert^2 \nonumber \\ &- \frac{\eta^2}{\kappa} \arctan\left(
\frac{\Delta_{c} - \frac{NU_0}{\pi} \int_0^{\pi} \romand x \vert \psi \vert^2
\cos^2(x)}{\kappa}\right). \label{eq:energyfunctional}
\end{align}
as can be verified by applying Hamilton's equation (\ref{eq:hamilton}). 
We shall refer to this functional as the reduced hamiltonian. The first term represents the kinetic energy. The second term is an
atom-light coupling dependent term
that can be interpreted as follows. The phase shift of the steady state light field inside the cavity relative to the pump laser is, from \eqnref{eq:stateqbm},
\begin{align}
\phi = \arctan{\frac{\Im(\alpha)}{\Re{(\alpha)}}} = \arctan\left(
\frac{\Delta_{c} - \frac{NU_0}{\pi} \langle \cos^2(x)
\rangle}{\kappa}\right) \ .
\end{align}
This allows us to rewrite the reduced hamiltonian as  
\begin{align}
E[\psi]= \frac{N}{\pi} \int_0^{\pi} \romand x \left\vert \frac{\romand \psi}{\romand x}
\right\vert^2 -I_{\mathrm{ph}} \phi \ ,
\end{align}
where $I_{\mathrm{ph}} = \eta^2/\kappa$ is the incident photon current from the pump laser. Note that this hamiltonian looks similar in form to
the
effective quantum two-mode hamiltonians obtained in the optomechanical limit in \cite{nagy09} and \cite{zhou09}, and also the ones describing
bistability in \cite{larson08}.

%%%%%%%%%%%%%%%%%%%%%%%%%%
\begin{figure*}[ht]
\includegraphics[width=\columnwidth]{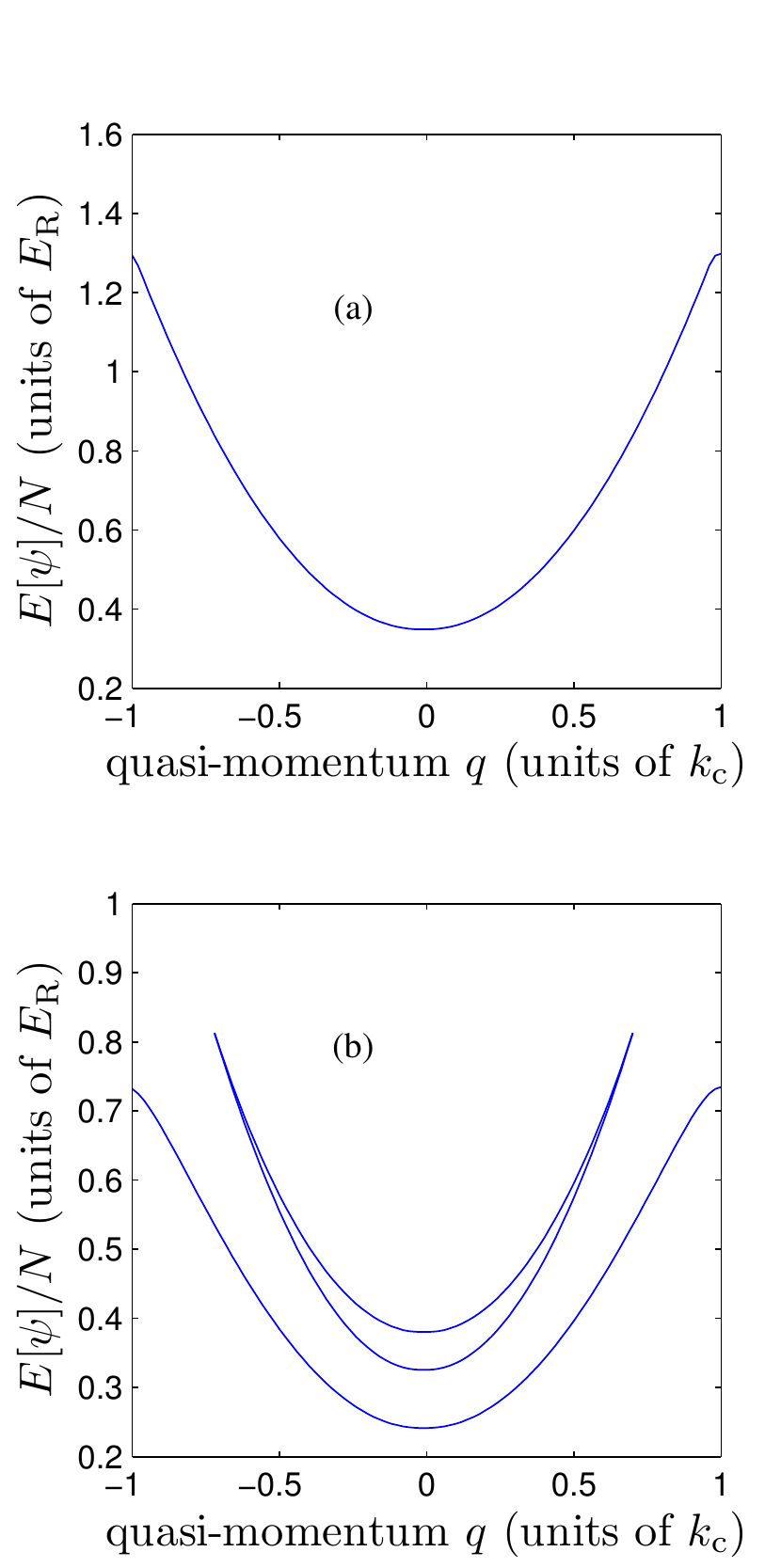}
\caption{Energy loops in the first band. The curves were obtained by extremizing the reduced hamiltonian (\ref{eq:energyfunctional}). For both
(a) and (b) the laser pumping $\eta = 909.9 \, \omega_{\mathrm{R}}$, the cavity decay $\kappa = 350 \, \omega_{\mathrm{R}}$, the atom-light
coupling $U_0 = \omega_{\mathrm{R}}$, and number of atoms $N=10^4$. In (a) the pump-cavity detuning was $\Delta_{c} =1350 \,
\omega_{\mathrm{R}}$, which gives no loops, and in  (b) it was $\Delta_{c} = 3140 \, \omega_{\mathrm{R}}$ which gives a loop symmetric about the
band center as shown. As explained in the text, the number of photons in the cavity and hence the lattice depth change with $q$. For example, in (a) at $q=0$
we have $n_{\mathrm{ph}} =0.06$ and at $q=1$ we have $n_{\mathrm{ph}} = 0.68$. In (b) at $q = 0$ we have for the lowest branch $n_{\mathrm{ph}} =
4.13$, for the middle branch $n_{\mathrm{ph}} = 0.28$, and for the upper branch $n_{\mathrm{ph}} = 2.4$. At $q=1$ we have $n_{\mathrm{ph}} =
1.08$. At the point where the middle and upper branches meet we have $n_{\mathrm{ph}} = 0.58$.}
\label{fig:pethickmethod1}
\end{figure*}
%%%%%%%%%%%%%%%%%%%%%%%%%%%%%%%%%

\section{Band structure}
\label{sec4}
From now on we specialize to Bloch wave solutions.  We begin by describing two
methods for calculating the Bloch waves and their energies. Agreement between the two methods will be demonstrated.

\subsection{Method 1: Energy extremization}\label{sec4a}
The first method, which adapts that detailed in \cite{machholm03} for a static optical lattice, hinges on the observation that the potential in
the Schr\"{o}dinger Eq.\ (\ref{eq:nonlineareqn}) is periodic with period $\pi$ (in dimensionless units). Despite the nonlinearity, the Bloch
theorem \cite{bloch28,zener34,bpv09} applies so that the eigenfunctions can be written as the product of a plane wave with wavenumber $q$, called the
quasi-momentum, and an amplitude $U_{q}(x)$ which is periodic with the period of the lattice
\begin{align}
 \psi_{q}(x) &= e^{iqx} U_{q}(x), \label{eq:blochansatz}\\
U_{q} (x+\pi) &= U_{q}(x) \ . \nonumber
\end{align}
For the linear problem, the energies $E(q)$ are arranged into bands separated by gaps. In the so-called reduced zone scheme for the band structure,
$q$ lies in the first Brillouin zone $- 1 \le q< 1$, and the band energies are folded back into the first Brillouin zone.  

The periodicity of $U_{q}(x)$ implies that it can be expanded as a Fourier series. The Bloch wave can therefore be written
\begin{align}
\psi_{q}(x) = e^{iqx}\displaystyle\sum_{n} a_{q,n} e^{i2nx}.
\end{align}
The notation for the $n^{\mathrm{th}}$ expansion coefficient $a_{q,n}$ reflects the fact that it depends on the chosen value of $q$. This
expansion can now be substituted into the reduced hamiltonian \eqnref{eq:energyfunctional}, and the resulting function numerically extremized
with respect to the parameters $a_{q,n}$, maintaining the normalization of $\psi_{q}(x)$ throughout the procedure. We take the parameters
$a_{q,n}$ to be real for the same reasoning as given in \cite{machholm03}. The Fourier series is terminated at some $n=R$, which is determined by
the convergence of the energy eigenvalues as $R$ is varied.

In \figref{fig:pethickmethod1} we plot the low lying part of energy spectrum $E[\psi_{q}]$ as a  function of quasi-momentum resulting from the extremization.  The values of
$\kappa$ and $N$ are very close to the values used in the experiment with rubidium atoms described in \cite{brennecke08}, and the other
parameters are chosen so as to exhibit the interesting behavior to be discussed in the rest of the paper. The two panels of
\figref{fig:pethickmethod1} differ only in the value of the pump-cavity detuning $\Delta_c$. \figref{fig:pethickmethod1}a shows a result familiar
from linear problems involving quantum particles in periodic potentials, but  \figref{fig:pethickmethod1}b shows a very different story: there
are two other branches that together form a looped structure that is a manifestation of the nonlinearity of the reduced hamiltonian. As will be
discussed more below, the loop shown in \figref{fig:pethickmethod1}b belongs to the first band (in particular, it does not correspond to the
second band because it only extends over part of the first Brillouin zone).  Looped structures have been found before for BECs in static
optical lattices \cite{machholm03,wu02}. We will come back to the similarities and differences between our results and \cite{machholm03,wu02} in
the next section.

It is important to appreciate that, by virtue of the nonlinearity of the problem, the lattice depth $n_{\mathrm{ph}} U_{0}$ is different at each
value of the quasi-momentum shown in \figref{fig:pethickmethod1}. Furthermore, the lattice depth is different for each of
the curves even at the same values of $q$ (except at degeneracies). In this sense, the band structure plots we display in this paper are non-standard because for static
lattices the depth of the lattice is fixed for all values of $q$. In order to obtain the lattice depth at each point on a curve
$E[\psi_{\mathrm{q}}]$ in \figref{fig:pethickmethod1}, one should take the wave function that extremizes the energy functional
(\ref{eq:energyfunctional}) at that point and enter it into the integral  $\langle \cos^2(x) \rangle$ appearing on the righthand side of
\eqnref{eq:selfconsphnum} for the photon number. This change in lattice depth with detuning is  reported in \figref{fig:nonlorent}, but
for the reader's convenience we have included some values at particular points $q$ in the caption of \figref{fig:pethickmethod1}.

The fact that, depending on the detuning $\Delta_{c}$, there are either one or more steady state photon numbers for the cavity reminds us of the
optical bistability observed in \cite{gupta07,brennecke08}. There, as the cavity driving laser detuning was swept through the resonance,
bistability was observed for certain pump strengths $\eta$. The bistability derives from quantum effects \cite{jonasfermi} in the sense that it
is due to changes in the atomic center-of-mass wave function. It is distinct from standard semi-classical optical bistability~\cite{gardiner}.
The connection between the loops in the atomic band structure and optical bistability will be examined in detail in Section \ref{sec5}. To complete
this section we look at an alternative method for determining the eigenfunctions of the effective hamiltonian which makes the connection with
bistability more transparent.

\subsection{Method 2: Self-consistency equation}\label{sec4b}
In the second method, the strategy is to solve \eqnref{eq:selfconsphnum} directly for the steady-state photon number. This equation contains
$n_{\mathrm{ph}}$ both explicitly on the left hand side and implicitly on the right hand side through the atomic wave function
$\psi_{q}(x,n_{\mathrm{ph}})$ appearing in the integrand of the integral 
\begin{align}
\langle\cos^2(x)\rangle = \frac{1}{\pi}  \int_0^{\pi} \cos^2(x) \vert
\psi_{q}(x,n_{\mathrm{ph}}) \vert^2 \romand x.
\label{eq:redefintionofcoupling}
\end{align}
The photon number is not a parameter set from the outside (e.g.\ by the pump laser) but must be solved for self-consistently. In our notation for
the wave function we have therefore included $n_{\mathrm{ph}}$ in the list of arguments rather than the list of parameters. The dependence of
$\psi_{q}(x,n_{\mathrm{ph}})$ upon the number of photons is because the depth of the lattice in which the atoms sit is given by $U_{0} \,
n_{\mathrm{ph}}$, as can be seen directly from the Schr\"{o}dinger equation (\ref{eq:atomeqn}). 
Therefore, \eqnref{eq:selfconsphnum} must be solved self-consistently for $n_{\mathrm{ph}}$, and we will often refer to it as the
``self-consistency equation''. As mentioned in the introduction, the physical mechanism that gives rise to the feedback between the atoms and the
light stems for the atom-light coupling, Eq.\ (\ref{eq:coupling}) [or Eq.\ (\ref{eq:redefintionofcoupling})]. The atoms' spatial distribution controls the phase shift suffered by the light
on traversing the cavity, and hence the cavity resonance condition, and therefore the amplitude of the light in the cavity. At the same time, the
amplitude of the light determines the depth of the lattice which influences the atomic wave function.

One class of solutions to the self-consistency problem is provided by the Mathieu functions \cite{a+s}. In fact, these provide \emph{exact}
solutions because the Schr\"{o}dinger equation (\ref{eq:atomeqn}) for a particle in a sinusoidal potential is nothing but the Mathieu equation.
Despite the fact that the amplitude of the sinusoidal potential in (\ref{eq:atomeqn}) itself depends on the solution
$\psi_{q}(x,n_{\mathrm{ph}})$ of the equation, this amplitude evaluates to a constant because $\psi_{q}(x,n_{\mathrm{ph}})$ appears under the
integral given by Eq.\ (\ref{eq:coupling}). All that is necessary is that the wave function that enters into the integral is the same as the wave
function appearing in the rest of the Schr\"{o}dinger equation. This is the case precisely when the self-consistency equation (\ref{eq:selfconsphnum}) is satisfied. This leads us to a very important point: by virtue of the self-consistency equation (\ref{eq:selfconsphnum}), the photon number $n_{\mathrm{ph}}$ is \emph{completely equivalent}, in the sense of the information it carries, to the wave function $\psi$. Said another way, the Mathieu functions are specified by only two quantities: the value of the quasi-momentum and the depth of the potential. Thus, once we set the quasi-momentum, which is a parameter, the wave function is entirely determined by $U_{0} n_{\mathrm{ph}}$, where $U_{0}$ is also a parameter.  We shall frequently take advantage of the equivalence between $\psi$ and $n_{\mathrm{ph}}$ in the rest of this paper because it allows us to replace the wave function by a single number $n_{\mathrm{ph}}$.

We shall denote the Mathieu functions by $\chi_{m,q,n_{\mathrm{ph}}}(x)$, where $m$ is the band index. They satisfy Mathieu's equation which in
our problem takes the form
\begin{eqnarray}
 \Bigg(\frac{\partial^{2}}{\partial x^{2}}  & & +  U_{0} n_{\mathrm{ph}}\cos^{2}(x)\Bigg) \chi_{m,q,n_{\mathrm{ph}}}(x) \nonumber \\
 = &&  \mu_{m,q,n_{\mathrm{ph}}} \chi_{m,q,n_{\mathrm{ph}}}(x) \ . \label{eq:MathieuEq}
\end{eqnarray}
Our notation for the Mathieu functions indicates the full parametric dependence with the exception of the atom-light coupling strength $U_{0}$.
Note that in the Mathieu functions we list $n_{\mathrm{ph}}$ as a parameter, like $q$, because that is the role it plays in the Mathieu equation.
We therefore have that 
\begin{align}
\psi_{m,q}(x,n_{\mathrm{ph}}) = \chi_{m,q,n_{\mathrm{ph}}}(x). \label{eq:mathieuform}
\end{align}
Mathieu's functions can, of course, be written in Bloch form. In this paper we focus on the first band and so we shall suppress the band index
from now on. We therefore have $\chi_{q,n_{\mathrm{ph}}}(x)=\exp(i q x) U_{q,n_{\mathrm{ph}}}(x)$. Substituting this Bloch form into the
time-dependent Schr\"{o}dinger equation \eqref{eq:atomeqn} as $\psi_{q}(x,n_{\mathrm{ph}},t)=\chi_{q,n_{\mathrm{ph}}}(x) \exp(-i \mu t) $ one
obtains
\begin{eqnarray}
 \left(\frac{\partial}{\partial x}+ i q\right)^{2}U_{q,n_{\mathrm{ph}}}(x)
&+ & U_{0} n_{\mathrm{ph}}\cos^{2}(x) U_{q,n_{\mathrm{ph}}}(x) \nonumber \\
= && \mu_{q,n_{\mathrm{ph}}} U_{q,n_{\mathrm{ph}}}(x). \label{eq:BlochEq}
\end{eqnarray}
This equation can either be solved numerically from scratch, or a package such as \textit{Mathematica} can be used which gives the Mathieu functions for each
value of $q$, and $U_{0} \, n_{\mathrm{ph}}$. For a particular choice of $q$, these Mathieu functions can now be used in \eqref{eq:selfconsphnum}
to find the value(s) of $n_{\mathrm{ph}}$ that give self-consistency.

There are two main differences between method 1 and method 2 described above. Firstly, method 1 is a variational calculation, whereas method 2
exploits the definition of the steady state photon number to obtain a single nonlinear integral equation (\ref{eq:selfconsphnum}) which must be
satisfied by the atomic wave function. Secondly, in method 2 we can explicitly set the band index whereas the variational wave function used in
method 1 is rather more general.
In spite of these differences, we find that the two methods are in complete agreement (to within numerical accuracy) for all the parameter ranges we were able to test (for very deep lattices method 1 becomes unworkable because a very large number of terms in the Fourier expansion are required). We have also checked that the two methods agree for higher bands. 

It may at first seem surprising that two such seemingly different methods are equivalent. We emphasize that both stem from the non-linear Schr\"{o}dinger equation (\ref{eq:nonlineareqn}), which is just Mathieu's equation with a potential depth which is set self-consistently. Method 1 minimizes an energy functional that is derived from this non-linear Schr\"{o}dinger equation. In principle, one could minimize ansatze other than the Bloch functions we use here (e.g.\ localized functions), but these would not then satisfy the original non-linear Schr\"{o}dinger equation (\ref{eq:nonlineareqn}) exactly.  Method 2 is based upon the observation that wave functions that satisfy the self-consistency equation (\ref{eq:selfconsphnum}) are precisely the required solutions of the non-linear Schr\"{o}dinger equation (\ref{eq:nonlineareqn}) providing we restrict ourselves to solutions which are Mathieu functions. Again, one could find other types of solution, but these would not satisfy Eq.\ (\ref{eq:nonlineareqn}).

An interesting question to ask is whether the nonlinearity of our problem mixes the linear bands, so that, for example the self-consistent first
band is a superposition of Mathieu functions from different bands. This is not what we find. Instead, the solutions we have found all correspond
to being from one or other band, but not a superposition. Method 2, in particular, allows us to explicitly set the band index and we are
therefore able to identify all three curves shown in  \figref{fig:pethickmethod1} as belonging to the first band (we have also checked that the
Mathieu functions corresponding to all three curves are orthogonal to the Mathieu functions for the second band for the same three lattice
depths). Actually, we do not find loops in higher bands for the parameter values considered in \figref{fig:pethickmethod1}. Although in this
paper we restrict our attention to the first band, we have found that we can have loops in the higher bands as well. The calculation of energy
dispersions using the self-consistent method is numerically less demanding and so
we will continue to use the latter for the remainder of this paper. In the next section we discuss optical bistability and its relation to loops in
the band structure.

\section{Bistability and Loops}
\label{sec5}

%%%%%%%%%%%%%%%
\begin{figure}
\includegraphics[width=\columnwidth]{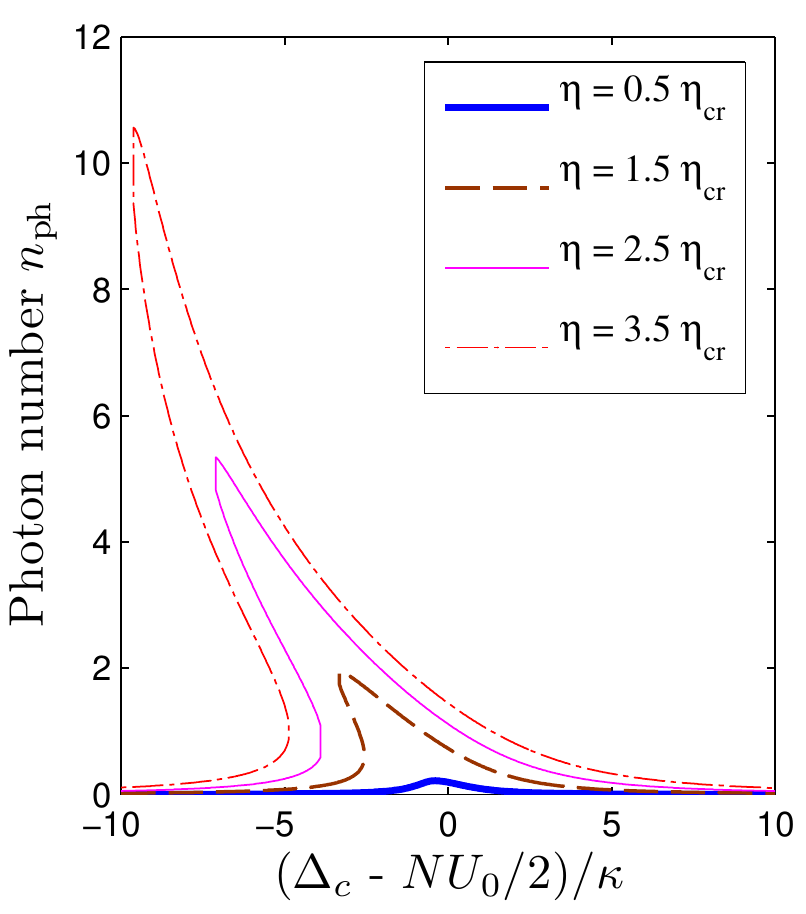}
\caption{(Color online) The steady state photon number inside the cavity as a function of detuning $\Delta_{c}$ for the parameters $\kappa = 350 \, \omega_{\mathrm{R}}$, $U_0 = \omega_{\mathrm{R}}$, $N=10^4$. Each curve is for a different value of the pump strength $\eta$:  thick blue $\eta = 0.5 \, \eta_\mathrm{cr}$, dashed brown $\eta = 1.5 \, \eta_{\mathrm{cr}}$, thin magenta $\eta = 2.5 \, \eta_{\mathrm{cr}}$, and dash-dotted red $\eta =3.5 \, \eta_{\mathrm{cr}}$.
As can be seen, as $\eta$ increases the lineshapes become more and more  asymmetric and fold over at the critical pump strength $\eta_{\mathrm{cr}}(q=0) \equiv \eta_{0} = 325 \, \omega_{\mathrm{R}}$. The atomic wave function corresponds
to the first band with $q=0$.}
\label{fig:nonlorent}
\end{figure}

\begin{figure}
\includegraphics[width=\columnwidth]{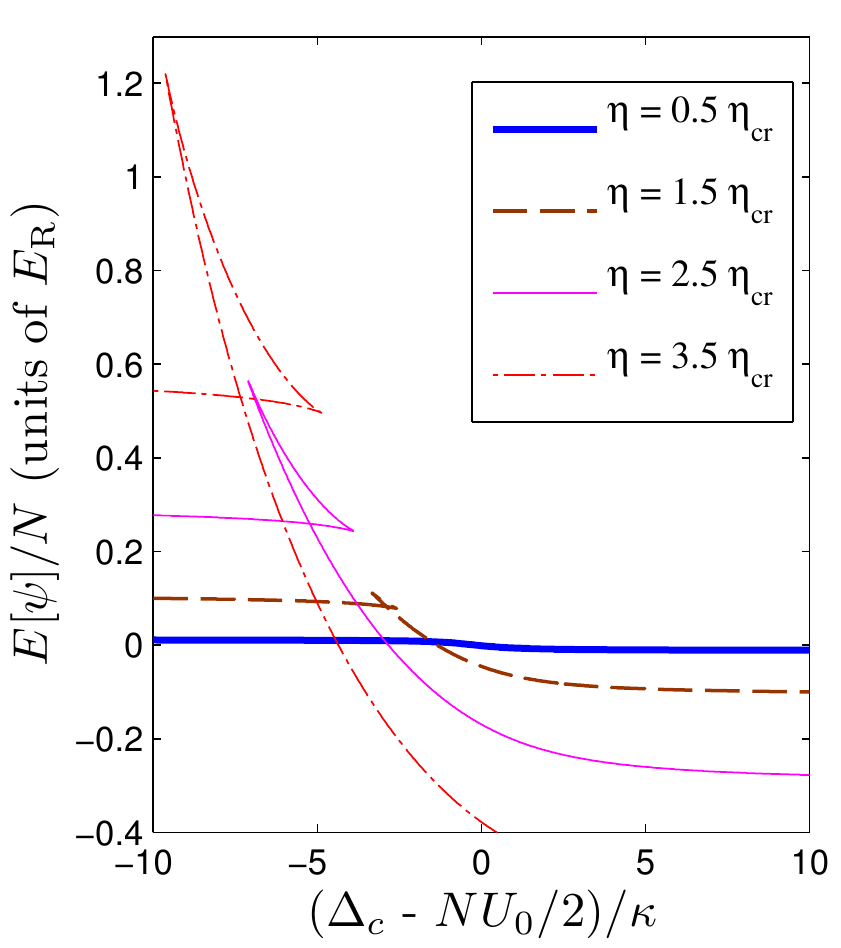}
\caption{(Color online) The energy given by the reduced hamiltonian (\ref{eq:energyfunctional}) as a function of detuning $\Delta_{c}$ for
the parameters $q=0$, $\kappa = 350 \, \omega_{\mathrm{R}}$, $U_0 =
\omega_{\mathrm{R}}$, $N=10^4$. Each curve is for a different value of the pump strength
$\eta$: thick blue $\eta = 0.5 \, \eta_\mathrm{cr}$, dashed brown $\eta = 1.5 \, \eta_{\mathrm{cr}}$, thin magenta $\eta = 2.5 \, \eta_{\mathrm{cr}}$, and dash-dotted red $\eta =3.5 \, \eta_{\mathrm{cr}}$. The critical pump strength is  $\eta_{0} = 325 \,
\omega_{\mathrm{R}}$. For $\eta>\eta_{0}$ swallowtail loops develop
corresponding to bistability. The loops grow in size as $\eta$ increases.}
\label{fig:nonlorenterg}
\end{figure}
%%%%%%%%%%%%%%%%

As mentioned in the Introduction, optical bistability was discovered in the ultracold atom experiments
\cite{gupta07} and \cite{brennecke08} via a hysteresis effect as the detuning of the pump laser was swept from above and below the cavity
resonance. This effect can be described theoretically by using the self-consistency equation to calculate the number of photons $n_{\mathrm{ph}}$
in the cavity at each value of the detuning (the number of cavity photons can be measured directly from the photon current transmitted through
the cavity which is given by  $n_{\mathrm{ph}} \kappa$). The results are plotted in \figref{fig:nonlorent} for different values of the pump
strength $\eta$.  In the absence of atoms the cavity lineshape is a lorentzian centered at $\Delta_{c} = 0$ with a full width at half maximum
$2\kappa$. At small pump intensity, the presence of the atoms shifts the center of the resonance by $NU_0\langle \cos^2(x) \rangle$ while the
shape is largely unaltered. But as $\eta$ is increased, the lineshape curve becomes asymmetric and eventually folds over when $\eta$ is above a critical
value $\eta_{\mathrm{cr}}(q=0) \equiv \eta_{0}$, indicating multiple solutions and hence bistability. $\eta_{\mathrm{cr}}(q)$ depends on the quasi-momentum and $\eta_{0}$ is defined as its value at $q=0$.

In the folded over region, only one the solutions (corresponding
either to the highest or the lowest photon number) is seen at a time, depending upon the direction of the sweep. The middle branch cannot be
accessed using this experimental protocol and is in any case unstable, as will be discussed at more length in Section \ref{sec10}. 

Note that in \figref{fig:nonlorent} we have chosen the quantum state of the atoms to be the $q=0$ Bloch state. In fact, this is the case for all
the plots in this section because that is closest to the situation in the experiments that have been conducted so far. One of the main points of this paper is essentially to examine the extra degree of freedom conferred by the quasi-momentum.  For atoms in ordinary optical lattices the quasi-momentum can be controlled by accelerating the lattice (rather than the atoms) via a phase chirp \cite{acc}. An accelerated lattice is hard to achieve inside a Fabry-Perot cavity, but if the atoms have a magnetic moment one can instead subject them to an inhomogeneous magnetic field so that a force is applied (or, of course, in a vertically oriented cavity the atoms will be accelerated by gravity).  As is well known from the theory of Bloch oscillations, under a constant force $F$ the quasi-momentum evolves according to the Bloch acceleration theorem \cite{bloch28,zener34} 
\begin{equation}
q(t)=q_{0}+\frac{Ft}{k_{c} E_{R}}
\label{eq:BO}
\end{equation}
where $q$ and $t$ are expressed in the dimensionless units given in Section \ref{sec2}, and $q_{0}$ is the quasi-momentum at time $t=0$. The Bloch acceleration theorem requires that the evolution be adiabatic, and the implications of this for intra-cavity optical lattices have been discussed in \cite{bpv09}, albeit without loops. The effect of a constant force is thus to sweep the system through the band at a constant rate and, in principle, any quasi-momentum can be achieved by switching off the magnetic field after an appropriate time delay. 

%%%%%%%%%%%%%%%%%%%
\begin{figure}
\includegraphics[width=\columnwidth]{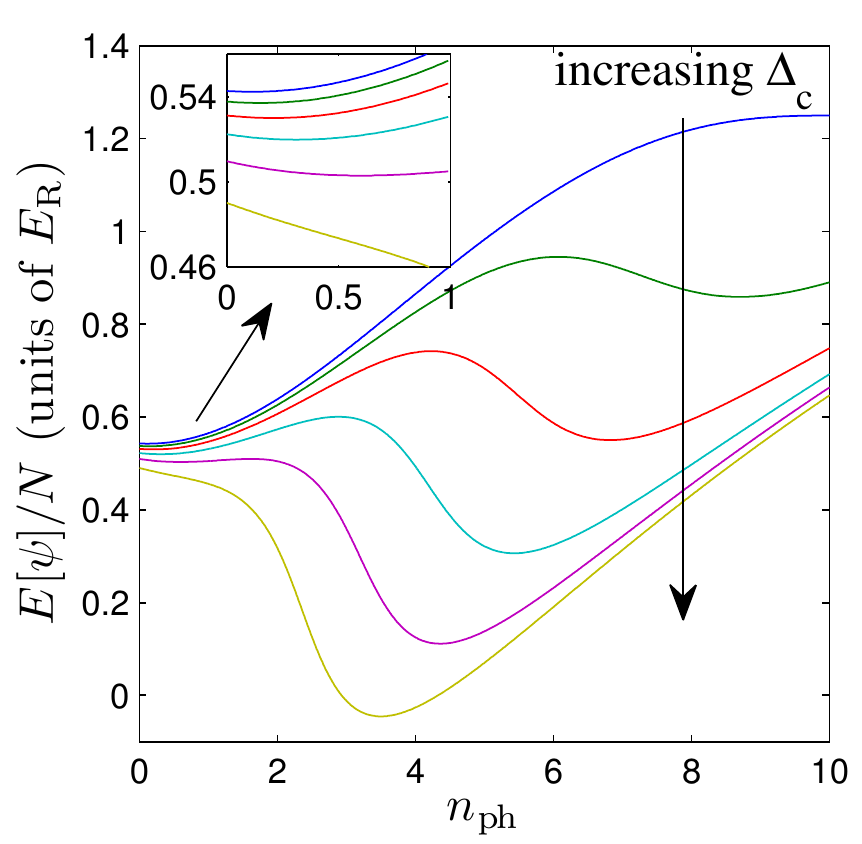}
\caption{(Color online) The double well structure of the energy of the reduced hamiltonian (\ref{eq:energyfunctional}) as a function of cavity photon number $n_{\mathrm{ph}}$.  Each curve is for a different value of the detuning $\Delta_{c}$. The values are
$\Delta_{c}=1600,2000,2400,2800,3200,3600 \, \omega_{\mathrm{R}}$. The arrow indicates how the curves evolve as $\Delta_c$ increases. The top most curve
(blue), and the bottom most curve (yellow), have only one minimum, whereas the rest of the curves have two minima (the inset shows a zoom-in of
the curves close to $n_{\mathrm{ph}} = 0$) indicating bistability. Consequently, bistability only occurs for a certain limited range of $\Delta_{c}$. The other parameters are $q=0$, $\kappa = 350 \, \omega_{\mathrm{R}}$, $U_0= \omega_{\mathrm{R}}$, $N=10^4$, $\eta = 3.5 \, \eta_{0}$, where
$\eta_{0} = 325 \, \omega_{\mathrm{R}}$.}
\label{fig:doublewell}
\end{figure}
%%%%%%%%%%%%%%%%%%%%%%

Figure \ref{fig:nonlorenterg} depicts the energy versus detuning curves corresponding to the photon number versus detuning curves shown in
\figref{fig:nonlorent}. In the bistable regime, the energy curves in Fig.\ \ref{fig:nonlorenterg} develop swallowtail loops. This can be understood in terms of the familiar
connection between bistability, hysteresis, and the change in the energy manifold described in detail by \cite{mueller02}. Consider one of the
curves in \figref{fig:nonlorenterg} where there is bistability, e.g.\ the curve with $\eta=3\eta_{0}$. For $\Delta_{c}$ values to the
left and right of the swallowtail loop, the energy functional \eqnref{eq:energyfunctional} has a single extremum corresponding to a particular
wave function $\psi_{q}(x,n_{\mathrm{ph}})$. In the bistable region, the energy functional has the structure of a double-well, furnishing three
extrema: two minima and one maximum, that give the three branches of the loop corresponding to three different wave functions. This double-well structure is shown in \figref{fig:doublewell} as a function of $n_{\mathrm{ph}}$ for
different values of detuning $\Delta_{c}$. Note that, as already observed above, the self-consistency equation (\ref{eq:selfconsphnum}) provides a direct mapping between the wave function and $n_{\mathrm{ph}}$.

Figures  \ref{fig:nonlorent}, \ref{fig:nonlorenterg}, and \ref{fig:doublewell} all show different aspects of hysteresis as $\Delta_{c}$ is swept either from above or below the cavity resonance.  It is enlightening to see how it arises in \figref{fig:doublewell}. If the detuning is swept from below the resonance then initially there is a single solution for $n_{\mathrm{ph}}$, given by the minimum in the reduced energy functional which occurs at the very left hand side of  \figref{fig:doublewell}, as best seen in the inset. When $\Delta_{c}$ is increased another solution appears at a larger value of $n_{\mathrm{ph}}$. However, this state of the system is not realized (for this direction of the detuning sweep), even when it becomes the global minimum, until the energy barrier between the two solutions vanishes and the left hand minimum disappears. The system then jumps to the new minimum at a larger value of $n_{\mathrm{ph}}$. The reverse happens when $\Delta_{c}$ is swept in the other direction. This hysteretic behavior is corroborated by Figs  \ref{fig:nonlorent} and \ref{fig:nonlorenterg}.

In the Section \ref{sec6}, a method for determining $\eta_{\mathrm{cr}}(q)$ is described. Generally, this requires a numerical computation, but for small values of the intracavity lattice depth an analytical expression can be worked out. It turns out that the dependence of
$\eta_{\mathrm{cr}}$ on the atomic state quasi-momentum can be used to explain the loops in the energy quasi-momentum plots (\figref{fig:pethickmethod1}). 

\section{Critical pump strength for bistability}\label{sec6}

%%%%%%%%%%%%%%%%%%%%%%%%%%%%%
\begin{figure}
\includegraphics[width=\columnwidth]{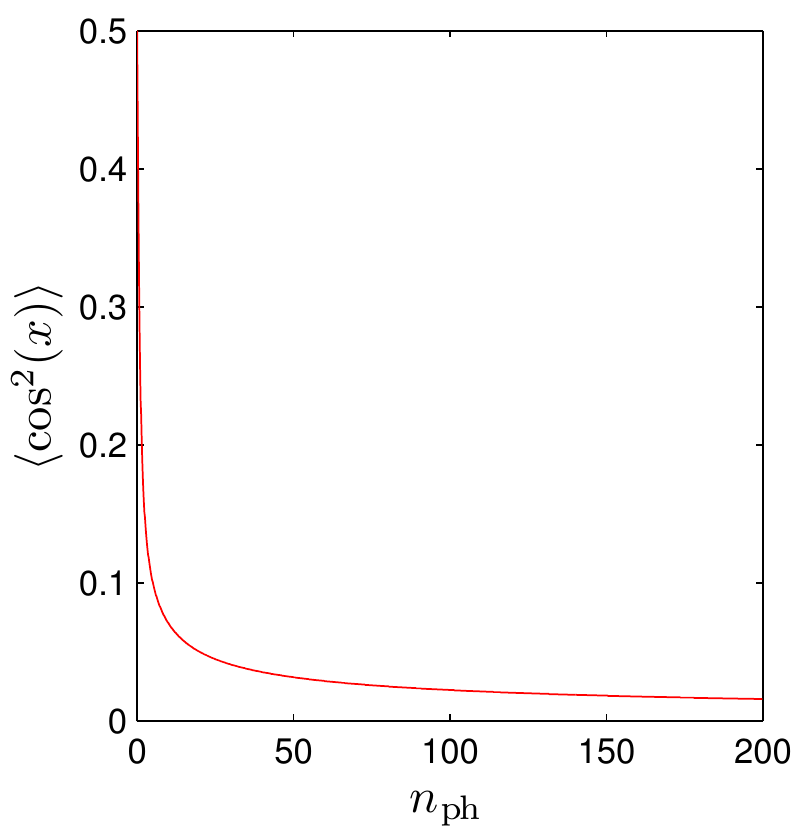}
\caption{(Color online) Plot of the atom-light overlap integral $f(n_{\mathrm{ph}},q) = \langle \cos^2(x) \rangle$, first defined in Eq.\ (\ref{eq:coupling}),
as a function of the cavity photon number $n_{\mathrm{ph}}$. The atomic wave function is taken to be the  
$q=0$ Bloch wave of the first band,  and the atom-light interaction is set at $U_0 = 5 \,
\omega_{\mathrm{R}}$. Note that the maximum value $f(n_{\mathrm{ph}},q)$ can take is one half, irrespective of the values of $U_0$ and $q$. As
$n_{\mathrm{ph}} \rightarrow \infty$ we find that $f(n_{\mathrm{ph}},q)\rightarrow 0$. }
\label{fig:cos2xfn}
\end{figure}
%%%%%%%%%%%%%%%%%%%%%%%%%%%%%%

\subsection{Conditions for bistability}

Returning to the cavity lineshape shown in Fig.\ \ref{fig:nonlorent}, we recall that as the pump strength $\eta$ is increased the steady state
photon number in the cavity can exhibit bistability for a certain range of detuning $\Delta_{c}$. Bistability first develops at a single value of
the detuning, which we denote by $\Delta_{c} = \Delta_0$. The critical pump strength at which this bistability at $\Delta_0$ occurs is $\eta_{\mathrm{cr}}(q)$, and in this section we want to calculate it. Let us first re-write the self-consistency equation (\ref{eq:selfconsrewrite}) as
\begin{align}
\frac{n_{\mathrm{ph}}}{n_{\mathrm{max}}} = \frac{1}{1+
\left(\frac{\Delta_{c}-NU_0 f(U_{0} n_{\mathrm{ph}},q) }{\kappa} \right)^2},
\label{eq:selfconsrewrite}
\end{align}
where $n_{\mathrm{max}} \equiv \eta^2/\kappa^2$ is the maximum number of the photons that can be in the cavity at steady state. In order to
reinforce the idea that the wave function and the number of photons are really equivalent quantities, we have replaced the notation for the
integral $\langle \cos^2(x) \rangle$ appearing in \eqref{eq:selfconsrewrite} by
\begin{equation}
f(U_{0} n_{\mathrm{ph}},q) \equiv \langle \cos^2(x) \rangle \ .
\end{equation}
This function is plotted in \figref{fig:cos2xfn} for blue detuning ($\Delta_{a}>0$) where we see that as the lattice gets deeper the atomic wave function adjusts to become more
localized on the lattice nodes and reduce the overlap between the light and the atoms. Furthermore, the steep gradient at shallow lattices
implies that the system is more sensitive, and so more nonlinear, at small photon numbers.

%%%%%%%%%
\begin{figure}
\includegraphics[width=\columnwidth]{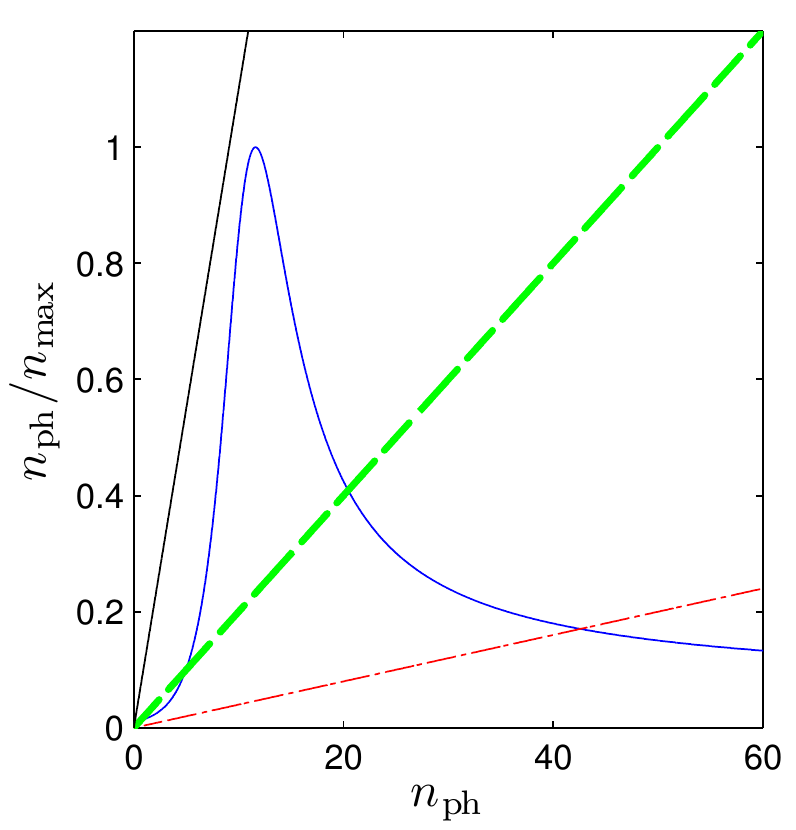}
\caption{(Color online) A graphical solution of the self-consistency equation in the form (\ref{eq:selfconsrewrite}). The blue curve represents
the right hand side 
of \eqnref{eq:selfconsrewrite} for typical values of the cavity
parameters ($q=0$). The red dash-dotted, green dashed and black straight lines represent the left hand side of the
equation plotted for different values of $n_{\mathrm{max}}$; they intersect the blue curve at one, three, and one points, respectively. The blue curve tends to a finite value at $n_{\mathrm{ph}}=0$ which is set by the fact that for $n_{\mathrm{ph}}=0$ we have $f=1/2$.}
\label{fig:graphsoln}
\end{figure}
%%%%%%%%%%%%%%%%

%%%%%%%%%
\begin{figure}
\includegraphics[width=\columnwidth]{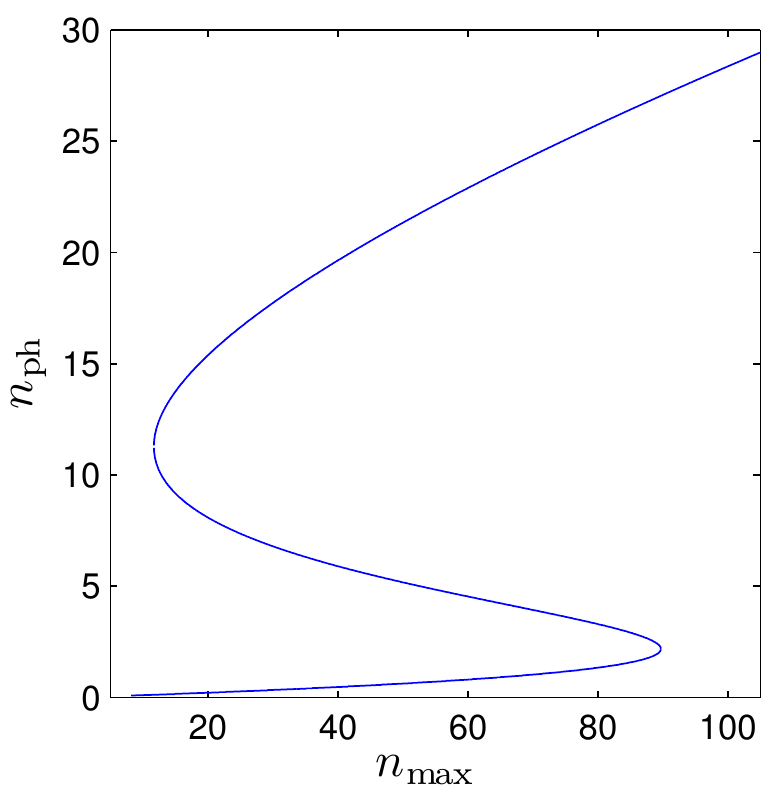}
\caption{(Color online) Input intensity vs output intensity for a bistable cavity system. In
this example the atomic wave function in the cavity is in the $q=0$ 
state and $\kappa = 350 \omega_{\mathrm{R}}$, $\Delta_{c} =1500 \omega_{\mathrm{R}}$, and $U_0 = \omega_{\mathrm{R}}$. The points where the curve
folds over are given by the solution of Eq.\ (\ref{eq:bistabcondn}).}
\label{fig:bistability}
\end{figure}
%%%%%%%%%%%%%%%%

It is instructive to solve \eqnref{eq:selfconsrewrite} graphically as the intersection between two functions of $n_{\mathrm{ph}}$, as shown in
\figref{fig:graphsoln}. The left hand side is a straight line  whose gradient is $1/n_{\mathrm{max}}$. For very small $n_{\mathrm{max}}$ the
gradient is very large and there is only one solution close to the origin. As $n_{\mathrm{max}}$ is increased the gradient is reduced and the
straight line just grazes the curve at a critical value of $n_{\mathrm{max}}$ at which there are now two solutions. Increasing $n_{\mathrm{max}}$
further, there is then a range of values of $n_{\mathrm{max}}$ for which there are three solutions. Finally, for large $n_{\mathrm{max}}$ there
is only one solution again. When three solutions exist for certain values of $n_{\mathrm{max}}$, the system becomes bistable and a plot of the
input intensity (proportional to $n_{\mathrm{max}}$) versus the output intensity (proportional to $n_{\mathrm{ph}}$) has the classic s-shaped
form shown in \figref{fig:bistability}. This picture suggests a convenient way to determine the conditions for bistability because the two points
where the curve turns over delimit the bistable region. These points satisfy $\partial n_{\mathrm{max}}/ \partial n_{\mathrm{ph}} = 0$, giving
\begin{align}
 \kappa^2 &+ \left(\Delta_{c} - NU_{0} f \right)^2
\nonumber \\
 &-2n_{\mathrm{ph}}\left ( \Delta_{c} -
NU_{0} f\right)NU_0 \frac{\partial f}{\partial n_{\mathrm{ph}}} = 0.
\label{eq:bistabcondn}
\end{align}
This equation can be solved numerically for $n_{\mathrm{ph}}$ for different values of $\Delta_{c}$, assuming that $\kappa$, $N$, and $U_0$ are
fixed.  As expected from Fig.\  \ref{fig:bistability} (see also Fig.\ \ref{fig:etadelbist}), depending on the value of $\Delta_{c}$, there are either zero, one, or two values of $n_{\mathrm{ph}}$ that satisfy
Eq.\ (\ref{eq:bistabcondn}). For large values of $\Delta_{c}$ there are no solutions. As $\Delta_\mathrm{c}$ is reduced, a single solution for
$n_{\mathrm{ph}}$ suddenly appears  at $\Delta_\mathrm{c}=\Delta_0$, which, by substituting this value of $n_{\mathrm{ph}}$ into the
self-consistency equation (\ref{eq:selfconsrewrite}), gives us $\eta_{\mathrm{cr}}(q)$. Reducing $\Delta_{c}$ further, this single solution
immediately branches into two solutions for $n_{\mathrm{ph}}$. Referring to \figref{fig:bistability}, we see that these two solutions for
$n_{\mathrm{ph}}$ define a range of values of $n_{\mathrm{max}}$, and hence also of $\eta$, where bistability occurs. We can find this range of
values of $\eta$ by inserting the two solutions for $n_{\mathrm{ph}}$ into the self-consistency equation to give us $\eta_1$ and $\eta_2$. When
$\eta_1<\eta<\eta_2$ bistability occurs. Note that the values of $\eta_{\mathrm{cr}}(q)$, $\eta_1$, and $\eta_2$, all depend on the state of the
atomic wave function and hence are different for different quasi-momenta. This dependence on quasi-momentum is what lies behind the existence of loops in the band structure.

\subsection{Critical pump strength in shallow lattices}

In the regime of shallow lattices, the bistability condition given by Eq.\ (\ref{eq:bistabcondn}) can be solved analytically. First consider the critical pump strength of the
$q=0$ case, for which we use the notation $\eta_{\mathrm{cr}}(q=0) \equiv \eta_{0}$. As described in \cite{brennecke08}, and as can be seen from \figref{fig:cos2xfn}, for small lattice depths we can linearize the
atom-light overlap integral as
\begin{align}
f(U_{0} n_{\mathrm{ph}},q=0) = \frac{1}{2} - \frac{U_0 n_{\mathrm{ph}}}{16} \label{eq:kerrnonlin}.
\end{align}
Substituting this into the self-consistency equation  (\ref{eq:selfconsphnum}), we obtain a cubic equation in $n_{\mathrm{ph}}$, which is
reminiscent of the classical Kerr nonlinearity in a medium with an intensity dependent refractive index~\cite{gardiner} (note that when we come
to numerically solve \eqnref{eq:bistabcondn} below in this section and in the rest of the paper, we are going beyond the classical Kerr effect).
The condition (\ref{eq:bistabcondn}) for bistability in this limit then reduces to the solution of a quadratic equation in $n_{\mathrm{ph}}$,
\begin{align}
3n_{\mathrm{ph}}^2N^2U_0^2
&-32n_{\mathrm{ph}}NU_0^2(NU_0-2\Delta_{c}) \nonumber \\ &+
64((NU_0-2\Delta_{c})^2+4\kappa^2)=0.
\end{align}
The vanishing of the discriminant of the above equation requires that
$\Delta_0 = NU_0/2-\sqrt{3}\kappa$, $\eta_{0}=
\sqrt{\frac{8\kappa^3}{3\sqrt{3}NU_0^2}}$, and $n_0 =
\frac{NU_0^2\sqrt{3}}{64\kappa}$. In the last expression, $n_{0}$ is the number of photons in the cavity at the critical point
$\Delta_{c}=\Delta_{0}$.

\subsection{Critical pump strength as a function of quasi-momentum}

\begin{figure}
\includegraphics[width=\columnwidth]{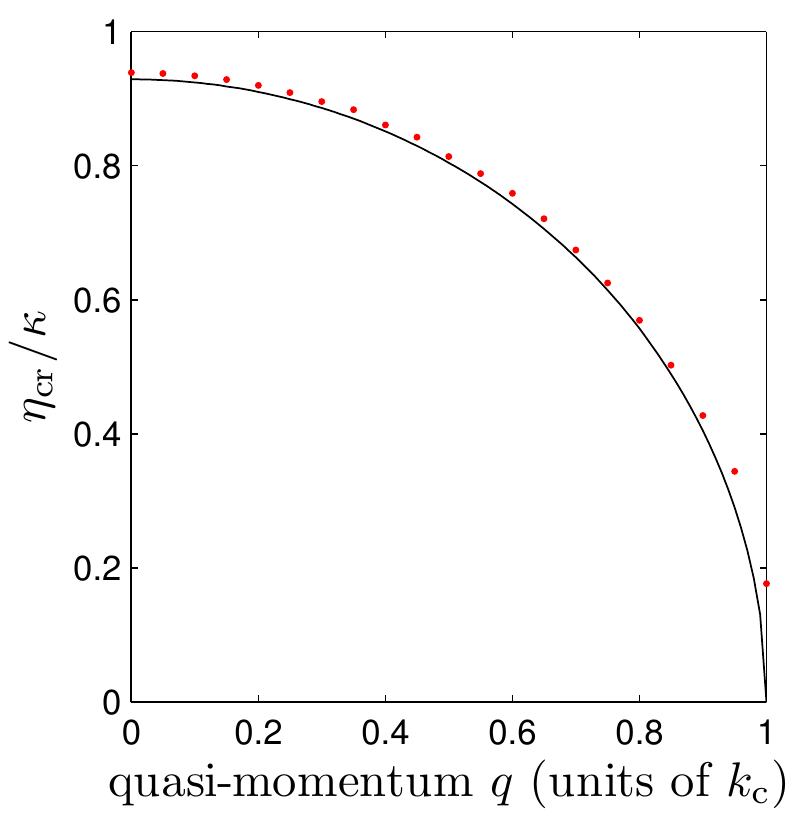}
\caption{(Color online) Comparison between exact numerical calculation (dots) and analytical estimate (line)  for the critical pump strength $\eta_{\mathrm{cr}}$ at which loops appear as a function of the quasi-momentum. The values of the parameters are
$U_0=\omega_R$, $N=10^4$, and $\kappa=350 \, \omega_R$. The analytical estimate is from \eqnref{eq:etacranalytic} which is accurate for small lattice
depths. Note that the agreement is good for quasi-momentum close to $q=0$.}
\label{fig:compecrnumanalytic}
\end{figure}

We now extend the above analysis for shallow lattices to include non-zero quasi-momentum. Expanding the atomic Bloch state in a Fourier series,
and assuming shallow lattice depths, we can truncate the series after three terms 
\begin{align}
\psi_{q}(x,n_{\mathrm{ph}},t) = e^{iqx}(c_0(t) + c_1(t) e^{i2x} + c_2(t) e^{-i2x}).
 \label{eq:wave}
\end{align}
In this state one can explicitly calculate $\langle \cos^2(x) \rangle$ as
\begin{align} 
\langle \cos^2(x) \rangle &= \frac{1}{2} + \frac{1}{2}(\Re{c_0c_2^{*}} + \Re{c_0
c_1^{*}}) \nonumber \\
&\equiv\frac{1}{2} + \frac{1}{2}(X(t)+Y(t)).
\end{align}
Rewriting the equations of motion (\ref{eq:atomeqn}) and (\ref{eq:lighteqn}) for the newly defined variables one finds
\begin{align}
\frac{\romand^2 X}{\romand t^2} + (4q+4)^2X + (q+1) U_0 \alpha^{*}\alpha |c_0|^2 = 0, \notag\\
\frac{\romand^2 Y}{\romand t^2} + (4q-4)^2Y - (q-1) U_0 \alpha^{*}\alpha |c_0|^2 = 0, \notag\\
\frac{\romand \alpha}{\romand t} = \left(i\Delta_{c}-i\frac{NU_0}{2}-i\frac{NU_0}{2}(X+Y) -\kappa\right)
\alpha + \eta. \notag
\end{align} 
The atomic state has therefore been mapped onto two coupled oscillators $X$ and $Y$. The oscillators are not coupled to each other directly, but
do interact through the light field $\alpha$ which acts as a driving term for both of them. The above equations resemble Eqns (3) and (4) of
\cite{brennecke08}, and introduce an analogy to optomechanics~\cite{kippenberg08}.   Solving these equations at steady state gives
\begin{align*}
\alpha &= \frac{\eta}{\kappa - i (\tilde{\Delta}_{\mathrm{c}} - NU_0/2(X+Y))},\\ 
(4q+4)^2 X + &\frac{(q+1)U_0\eta^2}{\kappa^2 + \left(\tilde{\Delta}_{\mathrm{c}}
- NU_0/2(X+Y) \right)^2} |c_0|^2 = 0,\\
(4q-4)^2 Y - &\frac{(q-1)U_0\eta^2}{\kappa^2 + \left(\tilde{\Delta}_{\mathrm{c}}
- NU_0/2(X+Y) \right)^2}|c_0|^2 = 0,
\end{align*}
where $\tilde{\Delta}_{\mathrm{c}} = \Delta_{c} - NU_0/2$. Combining the steady state solutions
into a single equation for the variable $p =
X+Y$, and assuming $\vert c_0\vert^2 \approx 1$ gives  
\begin{align}
p &= \frac{\bar{n}_{\mathrm{max}}}{1 + \left(\frac{\Delta_{c}}{\kappa}
-\frac{NU_0}{2\kappa}- \frac{NU_0 p}{2\kappa} \right)^2}, \label{eq:selfconsopt}
\end{align}
where $\bar{n}_{\mathrm{max}} = \frac{U_0 \eta^2}{8(q^2-1)\kappa^2}$. Comparing Eqns
(\ref{eq:selfconsopt}) and (\ref{eq:selfconsrewrite}) in the limit when
$f = 1/2 - U_0 n_{\mathrm{ph}} /16$, we finally obtain an expression for the critical pump
strength above which bistability occurs as a function of $q$ 
\begin{align}
\eta_{\mathrm{cr}} (q) = \sqrt{\frac{8\kappa^3 (1-q^2)}{3\sqrt{3}NU_0^2}} 
\label{eq:etacranalytic}
\end{align}
where we remind the reader that the frequencies in this expression are in units of the recoil frequency $\omega_{R}$.
This estimate for $\eta_{\mathrm{cr}}(q)$ is compared to the full numerical solution of
\eqnref{eq:bistabcondn} in \figref{fig:compecrnumanalytic}. The parameters are such that the maximum intracavity depth is only of the order of
one atomic recoil energy $E_{\mathrm{R}}$, and hence the approximation agrees well with the numerical calculation. The agreement deteriorates as
$q\rightarrow 1$. This is due to the fact that the above two mode approximation fails at $q=1$ because the coefficient $c_0$ in \eqnref{eq:wave}
is equal to zero at the edge of the Brillouin zone.

%%%%%%%%%
%%%%%%%%%%%%%%%%

Let us now connect the above results to the phenomenon of loops in the band structure. Because $\eta_{\mathrm{cr}}$, and also $\eta_1$, and
$\eta_2$ depend on the value of $q$, as we vary $q$  we expect that the conditions required to have bistability won't necessarily be met over the
entire Brillouin zone. That is, as $q$ is varied, $\eta$ may no longer lie in the range $\eta_1(q)<\eta<\eta_2(q)$. In that case, we expect any
additional solutions to form closed loops extending only over part of the Brillouin zone, rather than entire bands covering the whole Brillouin
zone. The dependence of $\eta_{\mathrm{cr}}$ upon $q$ given by Eq.\ (\ref{eq:etacranalytic}) suggests that for shallow lattices the loops will
form first at the edge of the Brillouin zone and then propagate inwards as $\eta$ is increased. Looking back at \figref{fig:pethickmethod1},
which shows the energy plotted as a function of quasi-momentum at a fixed value of $\eta$ and $\Delta_{c}$, we see a loop centered at $q=0$, but
which does not extend out to $q=1$, in apparent contradiction to what is predicted by Eq.\ (\ref{eq:etacranalytic}). This is because the lattice in 
\figref{fig:pethickmethod1} is too deep for Eq.\ (\ref{eq:etacranalytic}) to apply.  In the next section we shall examine how loops appear and
disappear and, in particular, we will see that loops are indeed born at the edges of the Brillouin zone.

\section{The birth and death of loops}
\label{sec7}

%%%%%%%%%
\begin{figure}
\includegraphics[height=0.75\paperheight]{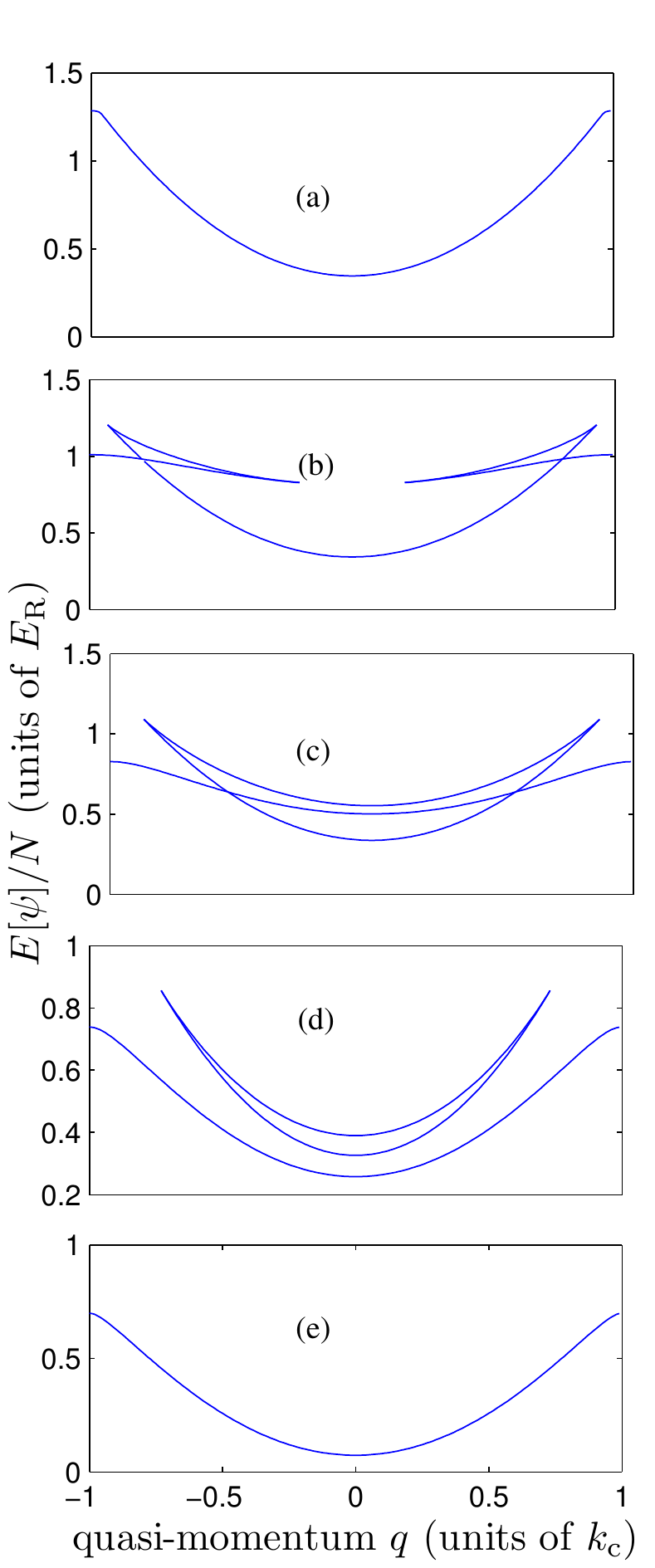}
\caption{(Color online) The birth and death of band structure loops as the laser-cavity detuning $\Delta_{c}$ is varied, for the case when the laser is blue-detuned from atomic resonance ($\Delta_{a}>0$). $\Delta_{c}$ increases as one goes from (a) to (e) as
follows: $1500 \, \omega_{\mathrm{R}}$, $2100 \, \omega_{\mathrm{R}}$, $2600 \,\omega_{\mathrm{R}}$, $3100 \, \omega_{\mathrm{R}}$ and $3600 \, \omega_{\mathrm{R}}$. The rest of the parameters are $\kappa = 350 \, \omega_{\mathrm{R}}$, $U_0 =
\omega_{\mathrm{R}}$, $N=10^4$, $\eta = 2.8 \, \eta_{0}$  and $\eta_{0} = 325 \, \omega_{\mathrm{R}}$.}
\label{fig:diffloops}
\end{figure}
%%%%%%%%%%%%%%%%

%%%%%%%%%
\begin{figure}
\includegraphics[height=0.75\paperheight]{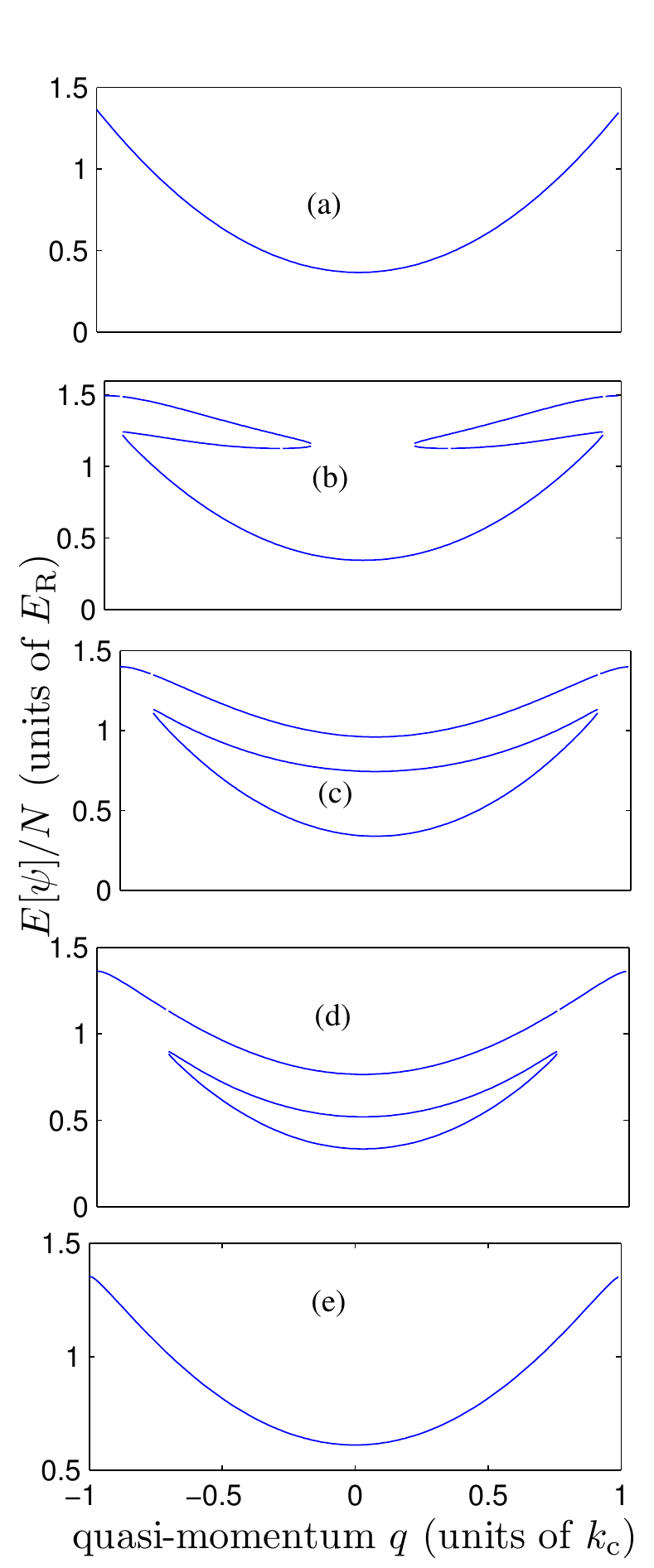}
\caption{(Color online) The birth and death of band structure loops as the laser-cavity detuning $\Delta_{c}$ is varied, for the case when the laser is red-detuned from atomic resonance ($\Delta_{a}<0$).
$\Delta_{c}$ increases as one goes from (a) to (e) as follows: $-8500 \, \omega_{\mathrm{R}}$, $-7900 \, \omega_{\mathrm{R}}$, $-7400
\,\omega_{\mathrm{R}}$, $-6900 \,
\omega_{\mathrm{R}}$ and $-6400 \, \omega_{\mathrm{R}}$. The rest of the parameters are $\kappa = 350 \, \omega_{\mathrm{R}}$, $U_0 = -
\omega_{\mathrm{R}}$, $N=10^4$, $\eta = 2.8 \, \eta_{0}$  and $\eta_{0} = 325 \, \omega_{\mathrm{R}}$.}
\label{fig:diffloopsU0neg}
\end{figure}
%%%%%%%%%%%%%%%%

%%%%%%%%%
\begin{figure*}
\includegraphics[width=\columnwidth]{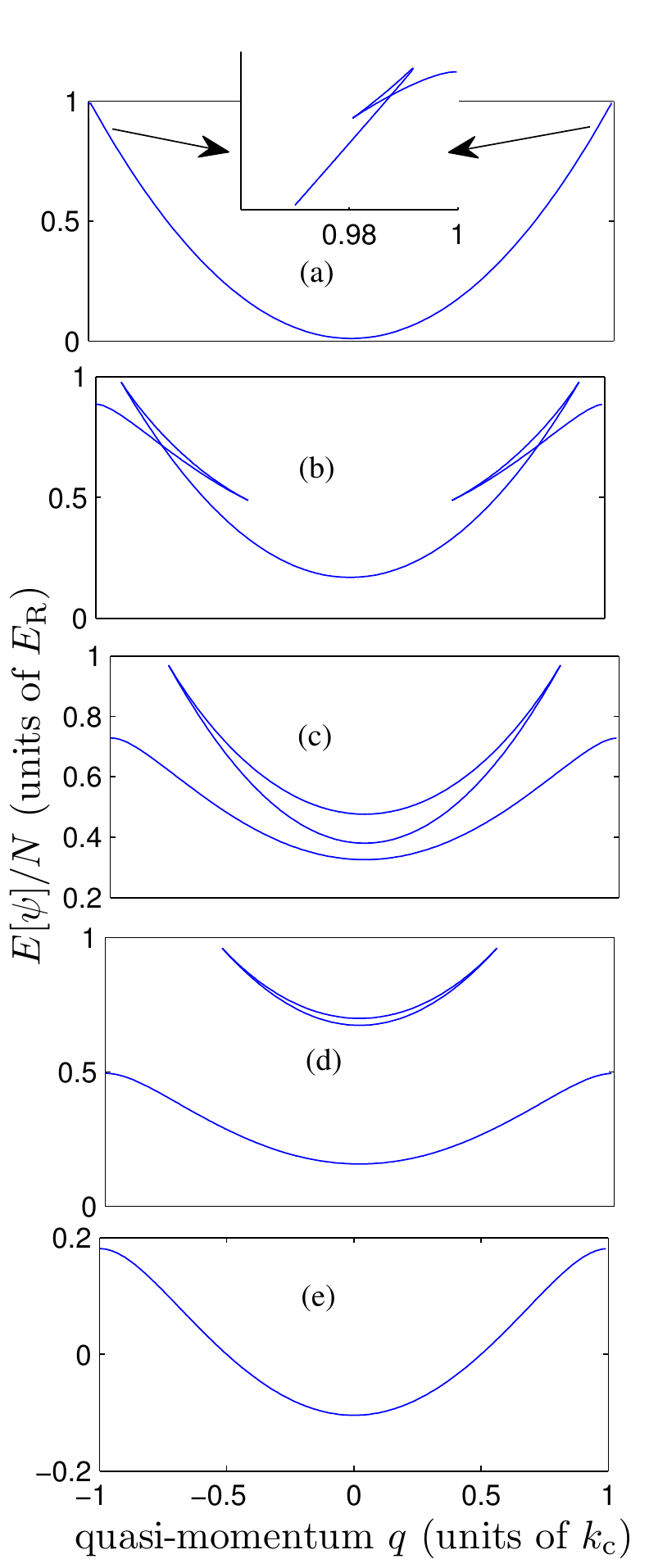}
\caption{(Color online) The birth and death of band structure loops as the pumping rate $\eta$ is varied. In this figure the detuning is held constant at $\Delta_c = 2900 \, \omega_R$. The value of $\eta$ increases from (a) to (e) as follows: $0.5 \eta_{0},2\eta_{0},3\eta_{0},4\eta_{0},5\eta_{0}$, where $\eta_{0} = 325 \, \omega_{R}$ as usual. The inset shows a zoom-in for $\eta = 0.5 \, \eta_{0}$, illustrating that as $\eta$ is increased, the loops are born at the edges of the Brillouin zone.}
\label{fig:diffloopseta}
\end{figure*}
%%%%%%%%%%%%%%%%

In this Section we examine how loops appear and disappear in the band structure as the detuning and pumping are varied. We have already seen in Sections \ref{sec4} and \ref{sec5} how multiple solutions and swallowtail loops develop as the detuning from the cavity resonance is varied, but this was for fixed quasi-momentum $q$. Here we include the quasi-momentum dependence. In \figref{fig:diffloops} we plot the evolution of
the loop structures that appear in the band structure as $\Delta_{c}$ is varied. The detuning increases from the top to the bottom panel. In the
plots, the pump strength is fixed at $\eta = 2.8 \eta_{0}$ and we see that for small detunings, when bistability initially sets in,
swallowtail shaped loops appear at the outer edges of the Brillouin zone. As the detuning is
increased the swallowtail loops from the two edges move closer and merge. Initially, the merged loop lies partly below the lowest band, but as the
detuning is further increased it moves up and separates from the lowest band. The loop then shrinks in size and vanishes.

One important point to notice is that the swallowtail loop in plot (b) of \figref{fig:diffloops} is qualitatively different from the ones obtained in \cite{machholm03,wu02} for an interacting BEC in an optical lattice because in our case the energy dispersion continues to have zero
slope at the band edge even when the loops have been formed. The nonlinearity in  \cite{machholm03,wu02}, which is due to interatomic interactions,  has quite a different form to that considered here. For example, the nonlinearity arising from interactions is spatially dependent due to the variation in density of a BEC in an optical lattice, whereas the nonlinearity considered here does not have a spatial dependence because it appears under an integral over space, see Eq.\ (\ref{eq:coupling}).

%%%%%%%%%
\begin{figure}
\includegraphics[width=\columnwidth]{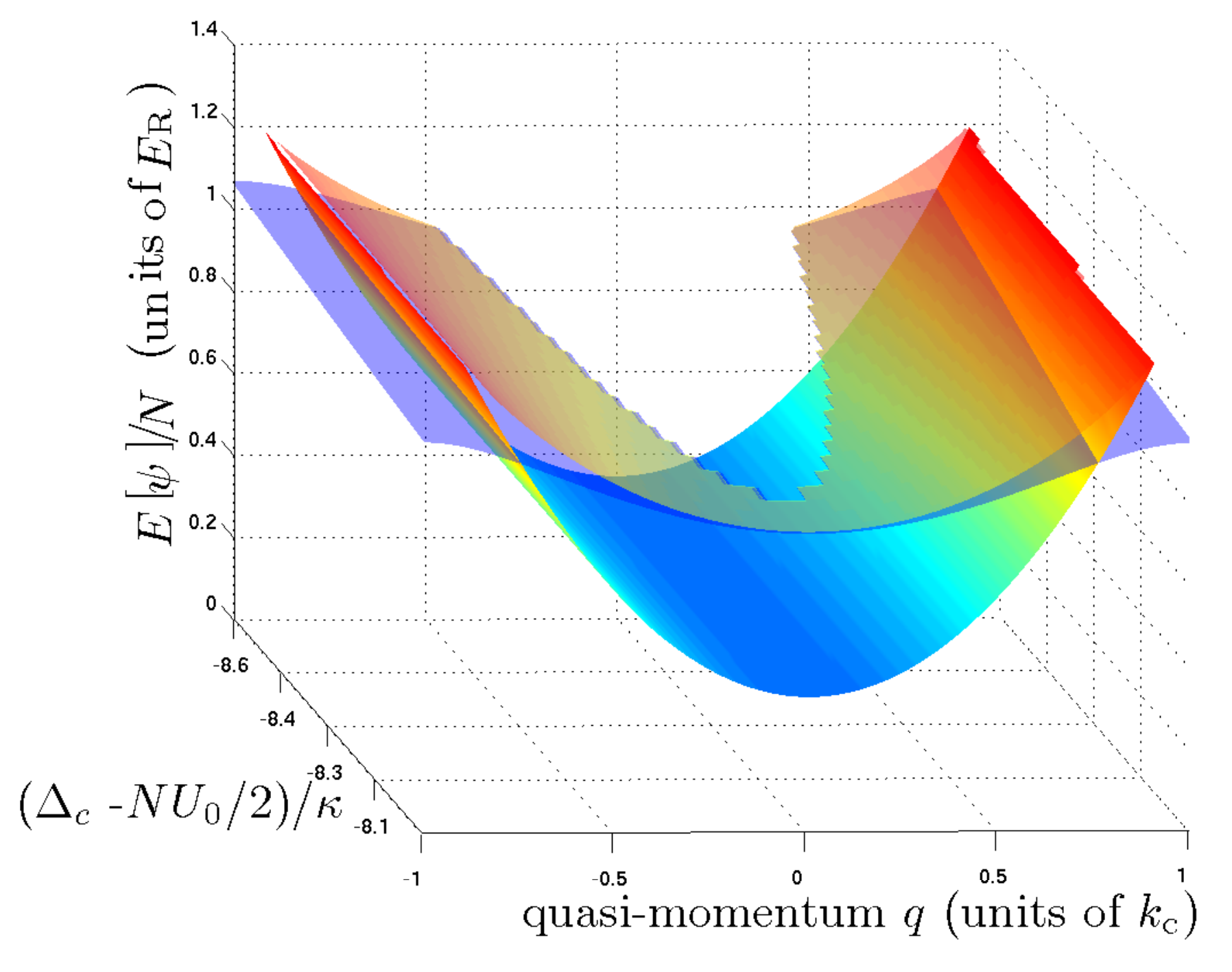}
\caption{(Color online) Energy as a function of quasi-momentum $q$ and detuning $\Delta_{c}$. At smaller values of $\Delta_{c}$, the swallowtail loops occur in pairs close to
the edges of the Brillouin zone at $q=\pm1$, and as the detuning is increased they propagate inwards and merge as shown in this plot. $\Delta_{c}$ increases out of the page. Parameters
are $\kappa = 350 \, \omega_{\mathrm{R}}$, $U_0 = \omega_{\mathrm{R}}$,
$N=10^4$, $\eta = 2.8 \, \eta_{0}$, and $\eta_{0} = 325 \,
\omega_{\mathrm{R}}$.
}
\label{fig:3dplot1}
\end{figure}
%%%%%%%%%
\begin{figure}
\includegraphics[width=\columnwidth]{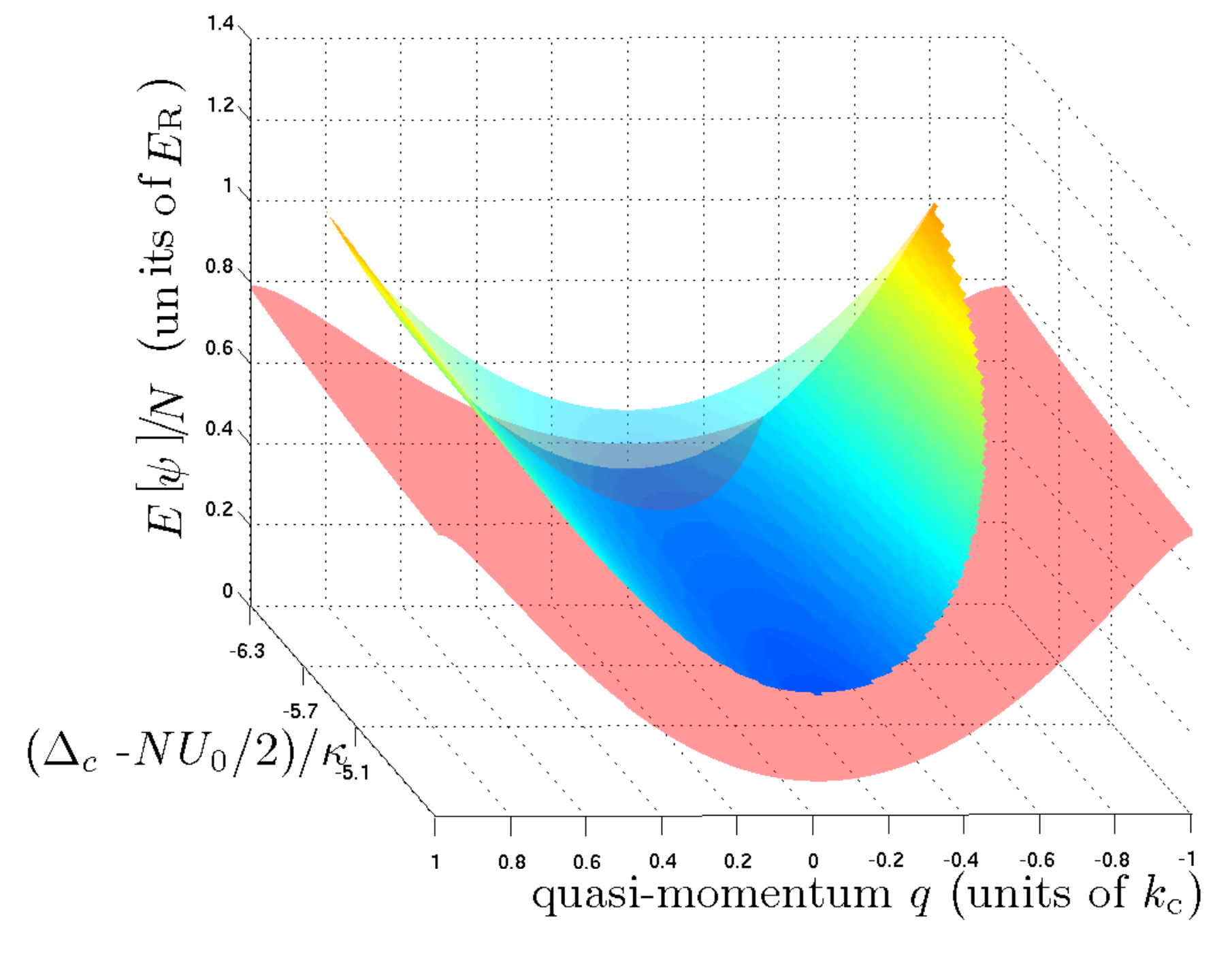}
\caption{(Color online) Energy as a function of quasi-momentum $q$ and detuning $\Delta_{c}$. $\Delta_{c}$ increases out of the page. For larger values of
$\Delta_{c}$, the band center loops move up in energy and do not touch
the lower band (shaded red in the plot). Eventually, for still higher values of
detuning, they shrink and disappear as shown in this plot. Parameters are $\kappa
= 350 \, \omega_{\mathrm{R}}$, $U_0 = \omega_{\mathrm{R}}$, $N=10^4$, $\eta =
2.8 \, \eta_{0}$, and $\eta_{0} = 325 \, \omega_{\mathrm{R}}$. 
}
\label{fig:3dplot2}
\end{figure}
%%%%%%%%%%%%%%%%
%%%%%%%%%%%%%%%%

In Fig.\ \ref{fig:diffloopsU0neg} we plot the same thing as \figref{fig:diffloops}, except that we have reversed the sign of the atom-light
coupling $U_{0}$ so that it is negative. Experimentally, this is the case when $\Delta_{a} <0$, i.e.\ the pump laser is red detuned from the
atomic resonance. Note that the effect of the sign flip upon the potential term $U_{0} n_{\mathrm{ph}} \cos^{2}(x)$ occuring in the atomic
Schr\"{o}dinger equation is equivalent to a spatial translation of $\pi/2$. This transforms the atom-light overlap integral (\ref{eq:coupling})
as 
\begin{equation}
\langle \cos^2(x) \rangle \rightarrow 1 - \langle \cos^2(x) \rangle
\end{equation}
where the left hand side refers to the $U_{0}<0$ case, and the right hand side to the $U_{0} >0$ case.
The self-consistency equation therefore becomes
\begin{align}
n_{\mathrm{ph}} &= \frac{\eta^2}{\kappa^2 + \left(\Delta_{c} +N \vert U_{0} \vert-
N \vert U_0 \vert f(\vert U_{0} \vert n_{\mathrm{ph}},q) \right)^2} \ .  \label{eq:selfconsphnumNegU0}
\end{align}

Figure \ref{fig:diffloopseta} plots the evolution of the band structure as the external laser pumping $\eta$ is varied, for fixed detuning $\Delta_{c}$. We see that as $\eta$ increased the loops first appear as swallowtails at the edges of the Brillouin zone (see inset in Fig.\ \ref{fig:diffloopseta}a), in agreement with the predictions of Eq.\ (\ref{eq:etacranalytic}).

Figures \ref{fig:3dplot1} and \ref{fig:3dplot2} are both 3D plots of the loops as functions of $\Delta_{c}$ and $q$, but each one covers a
different range of $\Delta_{c}$. In the first plot the merging of the swallowtail loops into the band center loops is shown. In the second plot,
as $\Delta_{c}$ increases the band center loops move upwards in energy and separate from the ground band, shrink in size and eventually
disappear.

As will be demonstrated in the next section, in certain parameter regimes the spectrum may qualitatively change compared to the above. In particular, we show that for some $q \neq 0$, and for sufficiently larger pump strength $\eta$, we can achieve tristability. Moreover, we show how this multistable behaviour can be understood in terms of catastrophe theory.

\section{Tristability}
\label{sec8}

%%%%%%%%%
\begin{figure}
\includegraphics[width=\columnwidth]{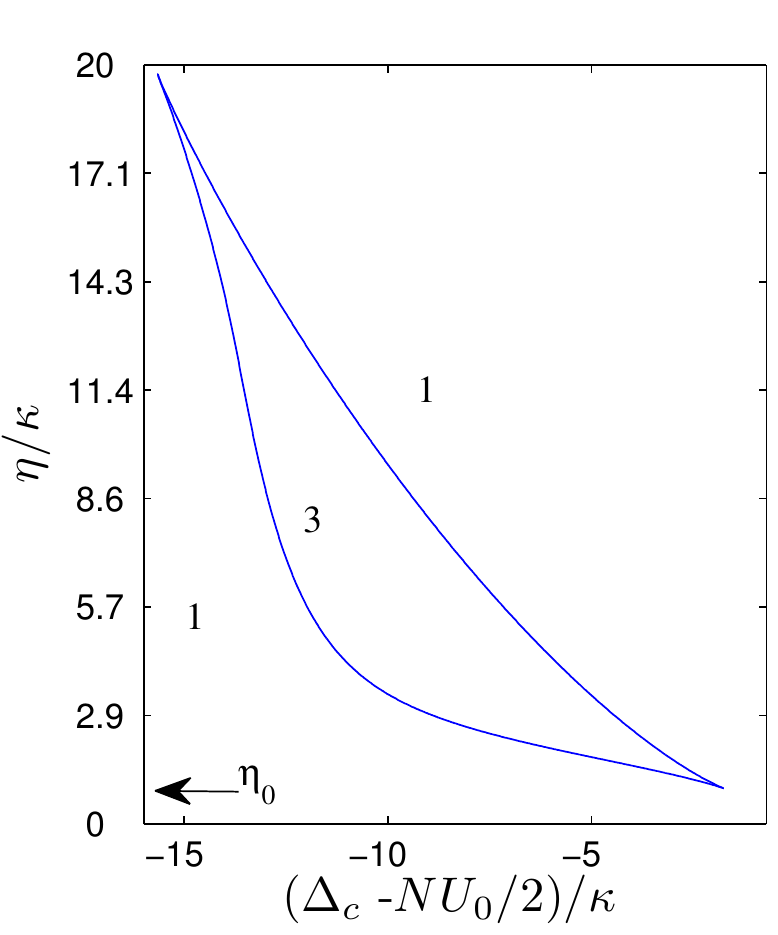}
\caption{(Color online) Bifurcation structure of the solutions to the self-consistency equation \eqnref{eq:selfconsphnum} in the $\{\eta,\Delta_{c}\}$ plane with $q=0$, $U_0 = \omega_{\mathrm{R}}$, $\kappa = 350 \, \omega_{\mathrm{R}}, N=10^4$. The numbers on the plot indicate the number of solutions that exist for the steady state photon number in the cavity. The critical value of the pumping $\eta_{0}=\eta_{\mathrm{cr}}(q=0)$ for bistability for these parameters is indicated by the arrow.  Inside the crescent shaped region the system supports three solutions (one unstable), i.e.\ it is bistable. }
\label{fig:etadelbist}
\end{figure}
%%%%%%%%%%%%%%%%

%%%%%%%%%
\begin{figure}
\includegraphics[width=\columnwidth]{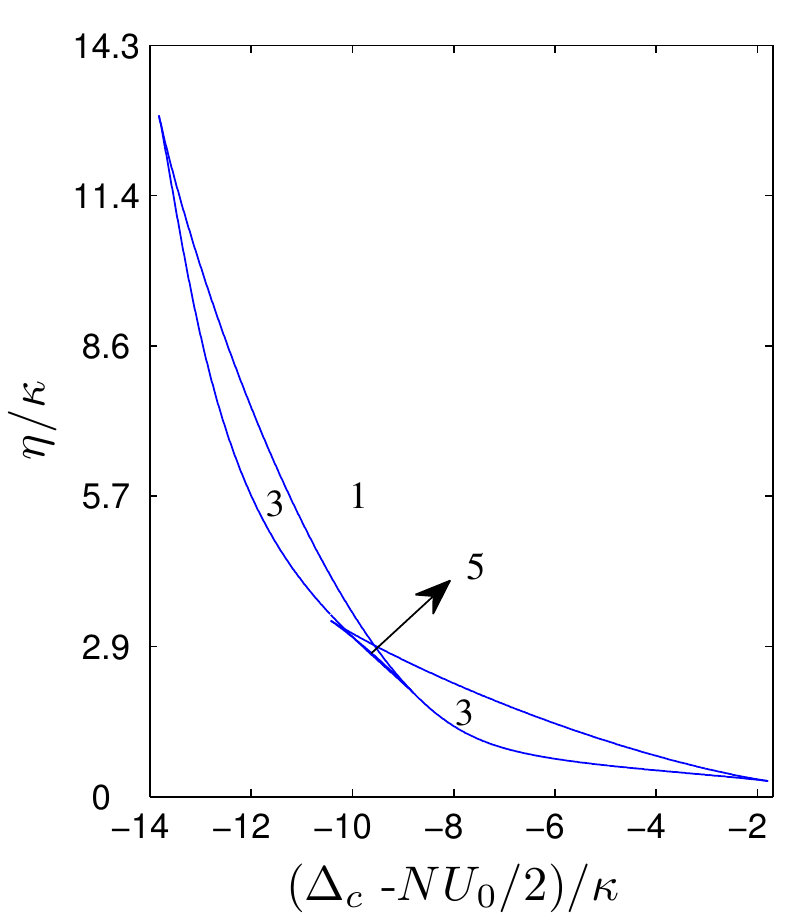}
\caption{(Color online) Bifurcation structure of the solutions to the self-consistency equation \eqnref{eq:selfconsphnum} in the $\{\eta,\Delta_{c}\}$ plane with $q=0.95$, $U_0 = \omega_{\mathrm{R}}$, $\kappa = 350 \, \omega_{\mathrm{R}}, N=10^4$. The numbers on the plot indicate the number of solutions that exist for the steady state photon number in the cavity. Inside the swallowtail shaped curve there are five solutions and hence multistability, see \figref{fig:multistab1}.}
\label{fig:etadelmultistab}
\end{figure}
%%%%%%%%%%%%%%%%

%%%%%%%%%
\begin{figure}
\includegraphics[width=\columnwidth]{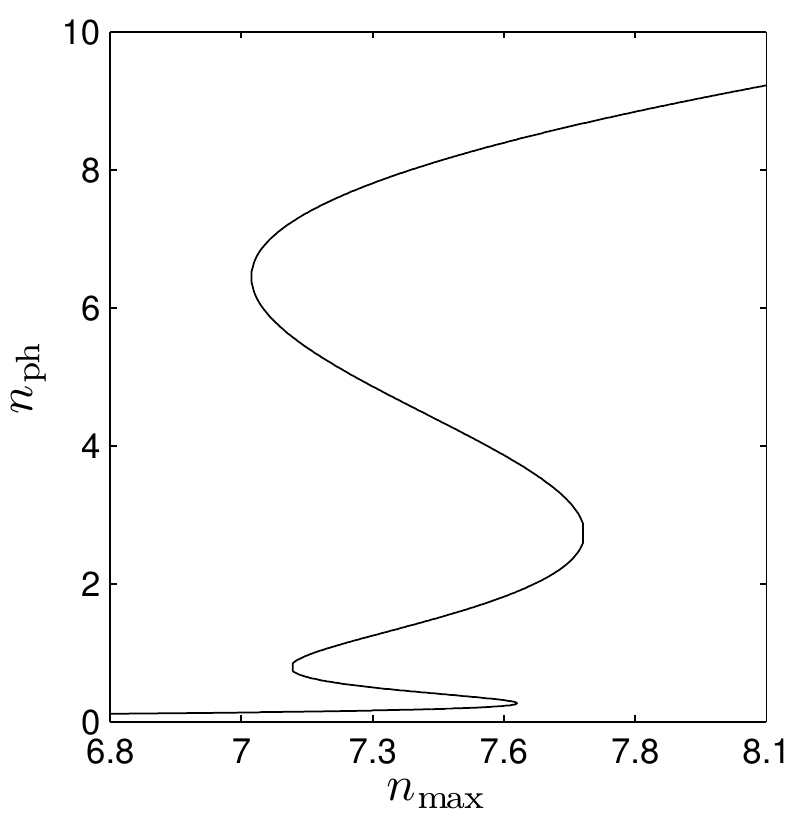}
\caption{Plot of multistable steady state photon number $n_{\mathrm{ph}}$ versus $n_{\mathrm{max}}$  (equal to $\eta^{2}/ \kappa^2$) for $\Delta_{c} = 1630 \, \omega_{\mathrm{R}}$ and $q=0.95$, $U_0 = \omega_{\mathrm{R}}$, $\kappa = 350 \, \omega_{\mathrm{R}}, N=10^4$.  The $\Delta_{c}$ value is chosen from
the region which supports five solutions in \figref{fig:etadelmultistab}.}
\label{fig:multistab1}
\end{figure}

%%%%%%%%%%%%%%%%

Thus far we have seen that there are regions of the parameter space $\{\Delta_{c},\eta,q,U_{0}\}$ where the self-consistency equation (\ref{eq:selfconsphnum}) applied to the first band admits either one or three solutions. In the latter case we have bistability. It is natural to ask whether there are other regions of parameter space where even more simultaneous solutions can occur? A recent paper discussing two-component BECs in a cavity has found regions of parameter space that support tristability \cite{dong11} and in this section we want to examine whether tristability is possible in the ordinary one-component case but with finite values of the quasi-momentum.

\begin{figure}
\includegraphics[width=\columnwidth]{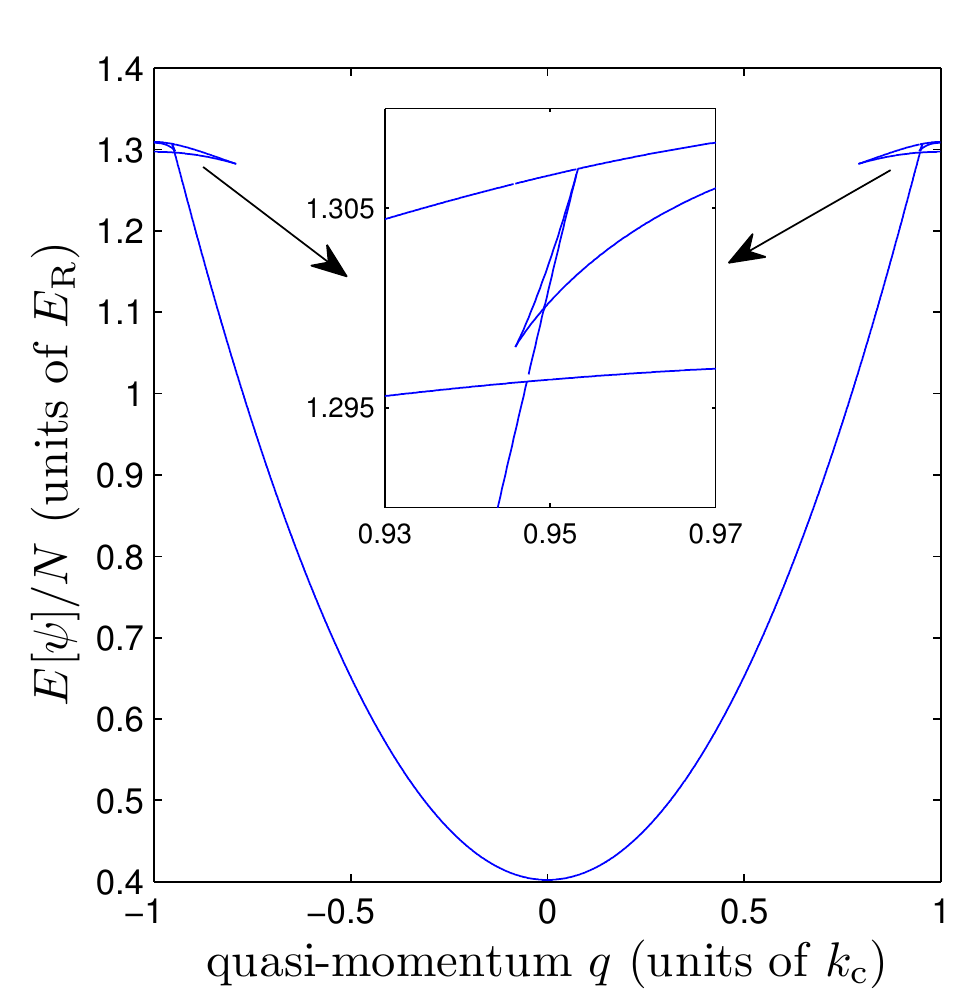}
\caption{(Color online) Plot of tristable band structure. The parameters are given by $\eta = 980 \, \omega_{R}$, $\Delta_c = 1640 \, \omega_{R}$, $\kappa = 350 \, \omega_{R}$, $N=10^4$, and $U_{0} =  \omega_{R}$. }
\label{fig:tristabqsp}
\end{figure}

In Fig.\ \ref{fig:etadelbist} we show the region of $\{ \Delta_{c},\eta \}$ space where bistability occurs. This plot is for fixed values of $U_{0}$ and $q$. In particular, for $q=0$ we at most find bistability. The crescent shape in Fig.\ \ref{fig:etadelbist} demarks the region with three solutions: the number of solutions changes by two as one crosses its boundary. The location of $\eta_{0}$, the smallest pump strength for which bistability occurs when $q=0$, is indicated by an arrow on the vertical axis in order to make contact with the discussion of Sec.\ \ref{sec5}. However, when we allow non-zero values of $q$ we find that we can indeed have regions of tristability, due to the presence of five solutions, which occur inside the swallowtail shaped region in Fig.~\ref{fig:etadelmultistab}, which is plotted for $q=0.95$. Fig.~\ref{fig:multistab1} plots the corresponding photon number versus laser pumping curve for a fixed value of $\Delta_{c}$, and illustrates how, as the laser pumping strength is changed, the system goes from one solution, to three solutions, to five solutions, back to three solutions and finally back to one solution again. This curve is calculated for a vertical slice through the parameter space shown in Fig.\ \ref{fig:etadelmultistab}. We give an example of the band structure when there is tristability in Fig.\ \ref{fig:tristabqsp}.

So far we have conducted a rather ad hoc exploration of the four-dimensional parameter space given by $\{\Delta_{c},\eta,q,U_{0}\}$. Furthermore, in the two-dimensional slices shown in Figures  \ref{fig:etadelbist} and \ref{fig:etadelmultistab} we have glimpsed snap shots of a rather complex looking structure of solutions. We are therefore led to ask whether there is any order in this complexity? Are the geometric structures seen in the plots random, or is there in fact an underlying structure that organizes them? In order to make progress with understanding multistability it would be useful to have a more systematic framework for analyzing our solutions and just such a framework is provided by catastrophe theory, which we now discuss. As will become clear, the structures we see in parameter space are not only generic, but are organized in a very particular, and therefore predictable, way.

%%%%%%%%%
%%%%%%%%%%%%%%%%

\section{Catastrophe theory analysis} 
\label{sec9}

\subsection{Overview of catastrophe theory}

%%%%%%%%%%%%%
\begin{figure}
\includegraphics[width=8.6cm,height=8.6cm]{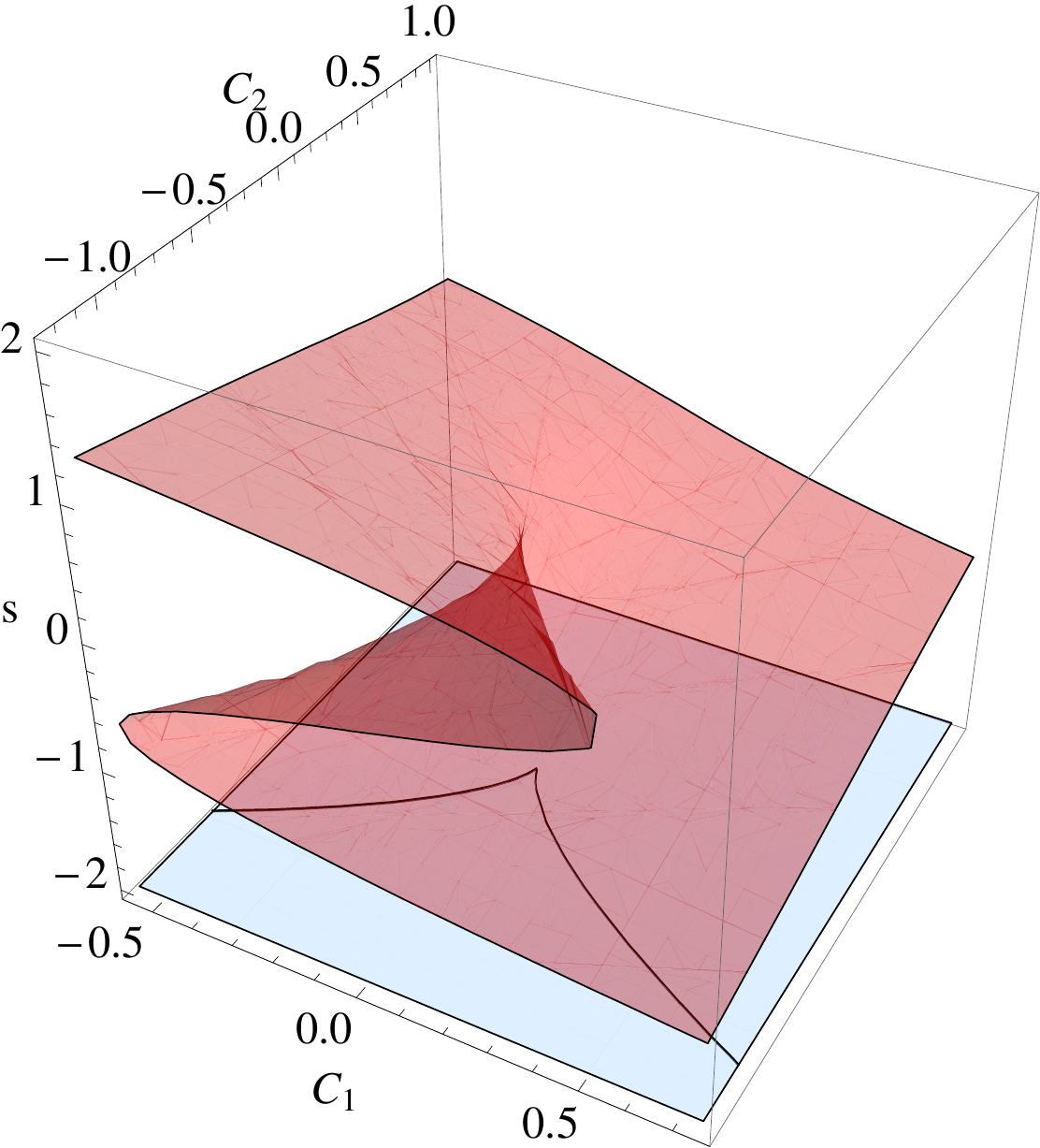}
\caption{(Color online) The cusp catastrophe is generated by the quartic potential function $\Phi$ given in Table \ref{tab:catastrophe}, which can be viewed as representing a double-well potential. The red folded-over surface plotted in this figure obeys the cubic state equation $\partial \Phi / \partial s = s^3+C_{2} s+C_{1} =0$, which gives the stationary solutions $s^{i}$ of $\Phi$. When $C_{2}<0$ there can be up to three stationary points, $s^{1}$, $s^{2}$ and $s^{3}$, for each value of $C_{1}$ and $C_{2}$. These points are the two minima and single maximum of the double-well potential. When $C_{2}>0$ there is only one stationary point $s$ corresponding to the minimum of a single well. A vertical slice through the figure such that $C_{1}=0$ gives a pitchfork bifurcation. The $\{C_{1},C_{2}\}$ plane forms the two dimensional control space where the cusp catastrophe itself lives,  and this is shown at the bottom of the figure. The cusp catastrophe is formed of two fold curves joined at a singular cusp point.  The cusp catastrophe demarks the region of control space that sustains three solutions for $s$, and so it is given by the projection of the folded-over part of the state surface onto the $\{C_{1},C_{2}\}$ plane. Crossing the fold lines from inside the cusp to outside it, two of the solutions (the maximum and one of the minima) annihilate. Mathematically, this is described by the potential function being stationary to the next higher order \cite{berry81}, i.e.\  $\partial^2 \Phi / \partial s^2 = 3 s^2+C_{2} =0$. Eliminating $s$ by combining this equation with the state equation gives the equation for the cusp catastrophe as $C_{1}=\pm \sqrt{-16 C_{2}^3/27}$. Right at the cusp point itself, which is given by the control space coordinates $C_1=C_2=0$, all three stationary points coalesce simultaneously to leave a single solution.}
\label{fig:cuspstd}
\end{figure}
%%%%%%%%%%%%%%%

Catastrophe theory is a branch of bifurcation theory that concerns the study of singularities of gradient maps \cite{berry81,nye99}. Examples of gradient maps abound in physics, for example Hamilton's principle of least action in mechanics, and Fermat's principle of least time in optics. In both theories the physical paths (rays) are given by the stationary points of a generating function $\Phi(\mathbf{s};\mathbf{C})$, which in mechanics is the action and in optics is the optical distance. In both cases the gradient map is obtained from a variational principle that takes the form
\begin{equation}
\frac{\partial \Phi(\mathbf{s};\mathbf{C})}{\partial s}=0 \ . 
\end{equation}
In catastrophe theory, this equation is sometimes referred to as the \emph{state equation}, the generating function $\Phi(\mathbf{s};\mathbf{C})$ is called the \emph{potential function} and the variables appearing in the potential function are considered to be of two basic types: the \emph{state variables} $\mathbf{s}=\{s_{1},s_{2},s_{3} \ldots\}$ and the \emph{control parameters} $\mathbf{C}=\{C_{1},C_{2},C_{3} \ldots\}$. The state variables parameterize all possible paths (not just the paths corresponding to the stationary points) and the control parameters determine the conditions imposed on the paths. For example, if we are interested in the path(s) which pass through the point $\{X,Y,Z\}$ in three dimensional space, then the coordinates $\{X,Y,Z\}$ form the control parameters. For a fixed set of control parameters, the potential function defines the height of a landscape with coordinates $\mathbf{s}$. The classical paths (rays) are then the stationary points $\mathbf{s^{i}}$ (labelled by the index $i$ if there are more than one) of this landscape, namely the mountain peaks, valley bottoms and saddles \cite{berry81}. 

The gradient map becomes singular when there is more than one stationary point for a given set of control parameters. In optics this is the phenomenon of focusing, because more than one ray passes through the same physical point $\mathbf{C}=\{X,Y,Z\}$, leading to a caustic. The caustic, or catastrophe, as it is known in catastrophe theory, lives in control space, which in the standard optics case is physical 3-dimensional space. As $\mathbf{C}$ is varied one can explore the structure of the caustic. This is  what was shown above in Figures  \ref{fig:etadelbist} and \ref{fig:etadelmultistab}, which are two dimensional slices through the parameter space $\{\Delta_{c},\eta,q,U_{0}\}$. The crescent and swallowtail shapes are the catastrophes, whose full structure only becomes apparent when viewed in four-dimensional parameter space. 

Catastrophes are points, lines, surfaces, and hypersurfaces in control space across which the number of solutions to the problem changes. Catastrophe theory classifies these catastrophes in terms of their codimension $K$, which is the difference between the dimension of the control space and the dimension of the catastrophe. For example, if we consider the two dimensional space shown in Figures \ref{fig:etadelbist} and \ref{fig:etadelmultistab}, we find ``fold'' curves ($K=1$) and ``cusp'' points ($K=2$). If we add a third dimension then we would find fold surfaces ($K=1$), cusp edges ($K=2$), and ``swallowtail'' points  ($K=3$). 

In order to make the foregoing discussion more concrete, consider the structure shown in Fig.\ \ref{fig:cuspstd} which illustrates the cusp catastrophe. The surface shown in the figure is the state surface $\partial \Phi / \partial s=0 $ plotted in a composite of control and state space. The control space is  two dimensional and is given by the $C_{1},C_{2}$ plane at the bottom of the figure. As listed in Table \ref{tab:catastrophe}, the cusp catastrophe is described by a quartic potential function which, by varying the control parameters $C_1$ and $C_2$, can be tuned between being a double or a single well potential. A prominent physical example of such quartic potential function is the thermodynamic potential in Landau's phenomenological theory  for continuous phase transitions \cite{landau+lifshitz}
\begin{equation}
\Phi(s;P,T,h)=\Phi_{0}+Bs^4+As^2+hs .\
\label{eq:landau}
\end{equation}
The order parameter $s$ for the phase transition can be identified as the state variable. The parameter $h$ describes an external field, and $A$ and $B$ are functions of pressure $P$ and temperature $T$. At first sight, it appears as though the Landau potential function contains three control parameters, and so does not correspond to the cusp catastrophe. However, it is easy to see that one of the parameters is redundant because the state equation can be written in terms of only two control parameters: $C_{1}=h/B$ and $C_{2}=A/B$. The Landau thermodynamic potential can therefore be seen to correspond to the cusp potential function.

In the present case of ultracold atoms in an optical cavity, the potential function is the reduced hamiltonian (\ref{eq:energyfunctional}), the state variable is the wave function $\psi$, and the control parameters are  $\{\Delta_{c},\eta,q,U_{0}\}$. We assume that the cavity decay rate $\kappa$ and number of atoms $N$ are constants that are unchanged throughout the analysis. The stationary Schr\"{o}dinger equation, obtained from the time-dependent Schr\"{o}dinger equation (\ref{eq:nonlineareqn}), is therefore the state equation which determines the allowed classical ``paths'' (rays) $\psi$. However, because we are interested here in solutions of the Bloch wave form, our ``paths'' $\psi$ are Mathieu functions labelled by the quasi-momentum and the depth of the optical lattice. The quasi-momentum $q$ is one of the control parameters, but the depth of the optical lattice is determined uniquely, via the self-consistency equation (\ref{eq:selfconsphnum}), by the number of photons in the cavity $n_{\mathrm{ph}}$. Therefore, we can choose to work with $n_{\mathrm{ph}}$ rather than $\psi$ as already discussed in Section \ref{sec4b}. The state equation which determines its stationary values is the self-consistency equation (\ref{eq:selfconsphnum}). 

\begin{table}
\caption{Standard forms of the cuspoid catastrophes}
\begin{tabular}{| l | l | l |}
\hline 
Name & $\Phi(s;C)$ & K \\
\hline 
Fold & $s^3/3 + C s$ & $1$ \\
Cusp & $s^4/4 + C_2 s^2/2 + C_1 s$&$2$\\
Swallowtail & $s^5/5 + C_3 s^3/3 + C_2 s^2/2 + C_1 s$& $3$\\
Butterfly& $s^6/6 + C_4 s^4/4 + C_3 s^3/3 + C_2 s^2/2 + C_1 s$& $4$\\
\hline 
\end{tabular}
\label{tab:catastrophe}
\end{table}

The purpose of formulating our problem in terms of catastrophe theory is that we can now take advantage of a very powerful theorem \cite{thom}. This states that there are only a strictly limited number of different forms which the potential function $\Phi(\mathbf{s};\mathbf{C})$ can take in the neighbourhood of a catastrophe. The first four are listed in Table \ref{tab:catastrophe}. Note that each of the standard forms is a polynomial in $s$, but is linear in the control parameters. In fact, for control spaces of dimension four or less there are only seven distinct structurally stable potential functions. Three of these require two state variables, whereas we only require one state variable, so that leaves the four so-called \emph{cuspoid} catastrophes, which are the ones listed in Table \ref{tab:catastrophe}. 

This remarkable result allows us to predict, at least qualitatively, the structures seen in Figures  \ref{fig:etadelbist} and \ref{fig:etadelmultistab} given only very rudimentary information such as the number of control parameters. However, the sting in the tail is that it is rare for the potential function that appears in any particular problem to already be in one of  the standard forms shown in Table \ref{tab:catastrophe}. Rather, it is generally necessary to perform various transformations upon the variables in order to manipulate the raw potential function into one of the standard forms. We already saw this for the rather simple case of the Landau theory discussed above, and we shall see below that this is also true for the problem of multistability in atom-cavity systems. 

From Table  \ref{tab:catastrophe}, we see that the butterfly potential function is the only one which gives up to five stationary solutions and has a four dimensional control space. We can therefore immediately say that our problem corresponds at least to a butterfly catastrophe because we have already found parameter regimes which give five solutions. This does not rule out the possibility of a higher catastrophe (the higher catastrophes contain the lower ones, as can be seen in Fig.\ \ref{fig:cuspstd} for the case of the cusp and the fold), but until we find regimes with a higher number of solutions (and, indeed, we have not) we shall work with the hypothesis that we are dealing with a butterfly catastrophe. We might therefore expect to find a very special point in parameter space where all five solutions merge into one (the ``butterfly point''). However, this requires us to be able to maneuver through parameter space in order to find this point.  The delicate issue of whether the four experimental parameters $\{\Delta_{c},\eta,q,U_{0}\}$ can be transformed into the butterfly's four linearly independent control parameters $\{C_{1},C_{2},C_{3},C_{4}\}$, and so allow us to fully explore the butterfly catastrophe, will be studied in Section \ref{sec:catastrophearbitrary} below. First, we begin with the simpler case of shallow lattices.

\subsection{Application of catastrophe theory to shallow lattices}
\label{sec:catastrophetheoryshallowlattice}

The starting point for our analysis will not be the potential function $\Phi$, which in our case is
the energy functional (\ref{eq:energyfunctional}) that explicitly depends on the atomic wave function $\psi_{q}(x,n_{\mathrm{ph}})$ and implicitly on the photon number $n_{\mathrm{ph}}$. Instead, we begin one step further along with the state equation which, as explained above, is provided by the self-consistency equation \eqnref{eq:selfconsphnum} for the photon number. This was also the approach adopted in \cite{agarwal79} in order to tackle bistability in traditional laser systems. The photon numbers that satisfy the state equation for given values of the control parameters form the set of stationary points of the potential function. For our state variable we shall actually choose 
\begin{equation}
v \equiv U_0 n_{\mathrm{ph}} \ .
\end{equation}
In terms of $v$ the state equation can be written
\begin{align}
 v + v\left(\Delta_{c} -  NU_0 f(v,q)\right)^2 - \eta^2 U_0 = 0. \label{eq:statefn}
\end{align}
where $\langle \cos^2(x) \rangle = f(v,q)$ is evaluated in a Bloch state $\psi_q(x,v)$, and the choice of  $v$ as the state variable is motivated by the dependence of the Bloch state on the product $U_0 n_{\mathrm{ph}}$, which is the depth of the optical lattice, see Eq.\ (\ref{eq:atomeqn}). In the above equation we have also rescaled all frequencies by $\kappa$ (i.e.\ we have divided throughout by $\kappa^2$ and set $\kappa = 1$). 

Equation (\ref{eq:statefn}) is not straightforward to analyze because we do not have a closed-form analytical expression for $f(v,q)$. Thus, we find ourselves in the common situation, as mentioned above, that it is not obvious which standard potential function, and hence which standard state equation, corresponds to our problem. However, we have already seen in Sec.~\ref{sec5} that when the depth of the optical lattice is small, we can perform a series expansion and obtain an approximate analytical expression for  $f(v,q)$. This is the approach we shall follow first and we will take up the general case in the next subsection. Specializing to the case of $q=0$, we use \eqnref{eq:kerrnonlin} to write $f(v,q=0) = 1/2-v/16$, and upon substitution into \eqnref{eq:statefn} this gives
\begin{align}
v^3 &+ b_1 v^2 + b_2 v + b_3 = 0, \label{eq:cuspnonred}\\
b_1 &= \frac{32 \Delta_{c}}{NU_0} - 16, \nonumber \\
b_2 &= \frac{64\left(4 + (NU_0-2\Delta_{c})^2\right)}{N^2U_0^2}, \nonumber\\
b_3 &= \frac{-256 \eta^2}{N^2 U_0^2}. \nonumber
\end{align}
The leading term in the above equation is cubic in the state variable $v$ (similar in that respect to the classical Kerr non-linearity \cite{meystre,gardiner}). It is therefore close to, but not yet identical with, the standard form of the state equation for a cusp 
\begin{align}
 s^3 + C_{2}s + C_{1} = 0.
\end{align}
Complete equivalence to the cusp can be achieved by removing the quadratic term from (\ref{eq:cuspnonred}) via the transformation $s = v + b_1/3$, leading to the following values for $C_{2}$ and $C_{1}$
\begin{align}
 C_{2} &= b_2 - \frac{1}{3} b_1^2, \label{eq:cuspmap1}\\
C_{1} &= b_3 - \frac{1}{3}b_1b_2 + \frac{2}{27} b_1^3. \label{eq:cuspmap2}
\end{align} 
Thus, we see that the canonical control parameters in the final mapping to the standard form are  complicated functions of the physical control parameters $\eta,\Delta_{c}$, and $U_0$. 

Let us compare the above prediction to the case illustrated in Fig.\ \ref{fig:etadelbist}. The 
figure shows the number of steady state solutions for the photon number in the $\{\eta,\Delta_{c}\}$ plane, for fixed values of $U_0$ and $q$. The figure was calculated for $q=0$, and the part of it close to the horizontal axis corresponds to low photon number, and so the shallow lattice theory outlined above applies in that region. We indeed find that the first derivative of the state equation vanishes at all points along the curves separating regions with one and three solutions (this is how the curves were computed), which are therefore fold curves, while at the point with $\eta = \eta_{0}$ we find the second derivative also vanishes, identifying it as a cusp point where all three solutions coalesce into a single solution. We therefore find that catastrophe theory correctly accounts for the structure seen in Fig.\ \ref{fig:etadelbist}.

A key point to note from the above analysis is that the underlying catastrophe that we identified had only two control parameters even though there were three ``experimental'' parameters that could be varied in the original statement of the physical problem (we set $q=0$). We met a similar situation for the Landau theory of continuous phase transitions discussed above. The question then arises, how do we identify the underlying catastrophe in cases where the transformation to standard form is hard to find? One way to proceed is via the defining character of each potential function in Table \ref{tab:catastrophe}, which is the highest derivative that vanishes at the most singular point. For the cusp the most singular point is $s=C_{1}=C_{2}=0$ where the first, second, and third derivatives of the potential function with respect to $s$ vanish. We shall take advantage of this defining character in order to tackle the general case of a lattice of arbitrary depth in the next subsection.

It is worth commenting on the role of the extra dimension in control space that was present in the original statement of the shallow lattice problem, as given by Eq.\ (\ref{eq:cuspnonred}). It is easy to imagine that, given a basic catastrophe, we can always embed it in a  control space of higher dimension without fundamentally changing the catastrophe providing we extend it into the extra dimension in a trivial way. For example, given the cusp catastrophe that has a minimal control space which is two-dimensional, as shown in Fig.\ \ref{fig:cuspstd}, we can always add a third dimension such that the cusp becomes a structure rather like a tent, with the ridge pole being a cusp edge (a continuous line of cusps).  The existence of the line of cusps can be inferred from the shift of variables  $s = v+b_1/3$ that was performed: the location of the highest singularity is parameterized by the parameter $b_1$ which is the extra control parameter.

\subsection{Application of catastrophe theory to lattices of arbitrary depth}
\label{sec:catastrophearbitrary}

Lifting the restriction of small photon number, we define the notation $\mathcal{G}(v;\Delta_{c},\eta, q, U_0)$ for the left hand side of the state equation (\ref{eq:statefn})
\begin{align}
 \mathcal{G}(v;\Delta_{c},\eta, q, U_0) \equiv v &+ v\left(\Delta_{c}  -  NU_0 f(v,q)\right)^2\nonumber \\& - \eta^2 U_0 = 0.
\label{eq:statefnre}
\end{align}
The fact that we have four experimental parameters holds out the possibility that the above state equation is that of a butterfly catastrophe (see Table~\ref{tab:catastrophe}). Furthermore, because we have already discovered regimes with five solutions,  we must have at least a butterfly. However, from our experience with the shallow lattice case,  we know that we may not be able to fully explore the catastrophe if some of the control parameters are trivial. We shall therefore investigate whether we can find a singular point (the butterfly point) where the fifth and all lower derivatives of the potential function with respect to $v$ simultaneously vanish  (i.e.\ the fourth and lower order derivatives of $\mathcal{G}(v;\Delta_{c},\eta, q, U_0)$ simultaneously vanish) and \eqnref{eq:statefnre} is also satisfied.  

%%%%%%%%%%%%%%%%%%%%%%%%%%%%%%%%%%%%

\begin{figure}
\includegraphics[width=\columnwidth]{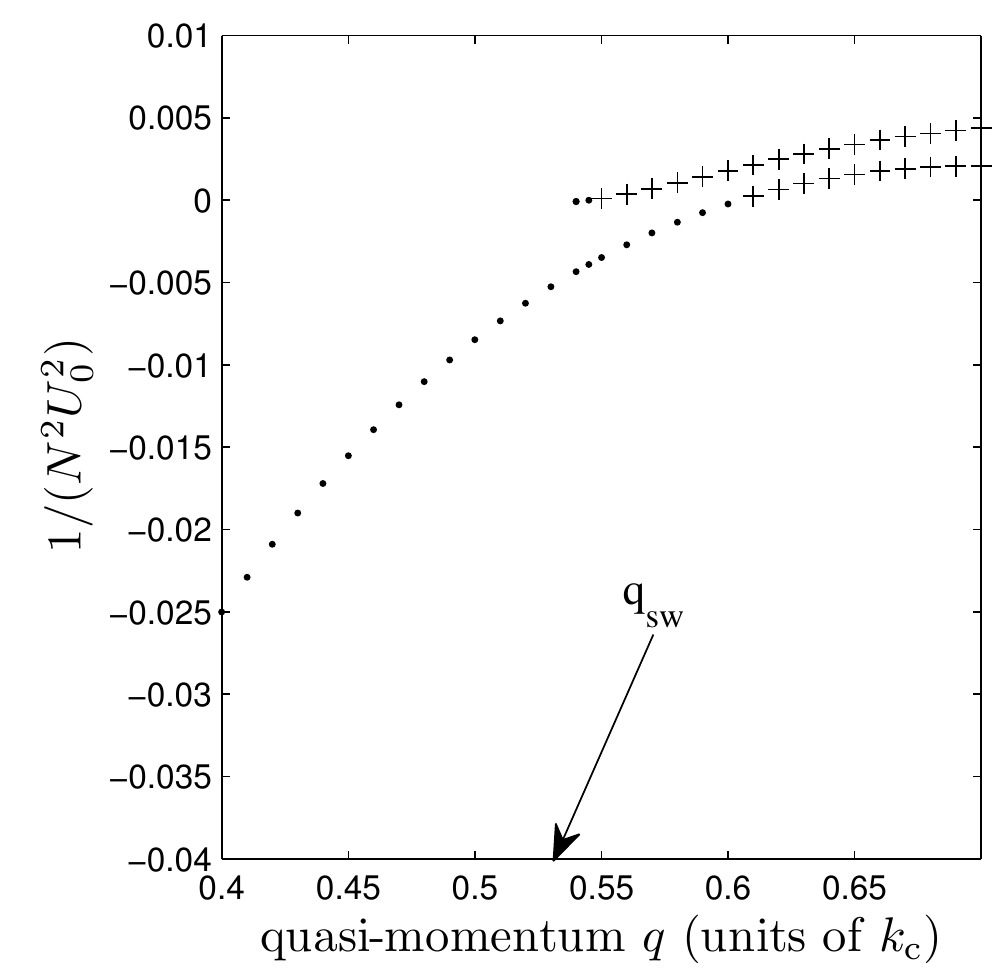} 
\caption{A plot showing the values of $1/(NU_0)^2$ obtained from the simultaneous solution of Eqns (\ref{eq:firstder})--(\ref{eq:thder}) for different values of quasi-momentum. These equations
give the first three derivatives of the state equation (\ref{eq:statefnre}), and hence correspond to swallowtail points. Only the crosses satisfy $1/(NU_0)^2 >0$ and occur only for $q>q_{\mathrm{sw}}$. }
\label{fig:u0zerocross}
\end{figure}

\begin{figure}
\includegraphics[width=\columnwidth]{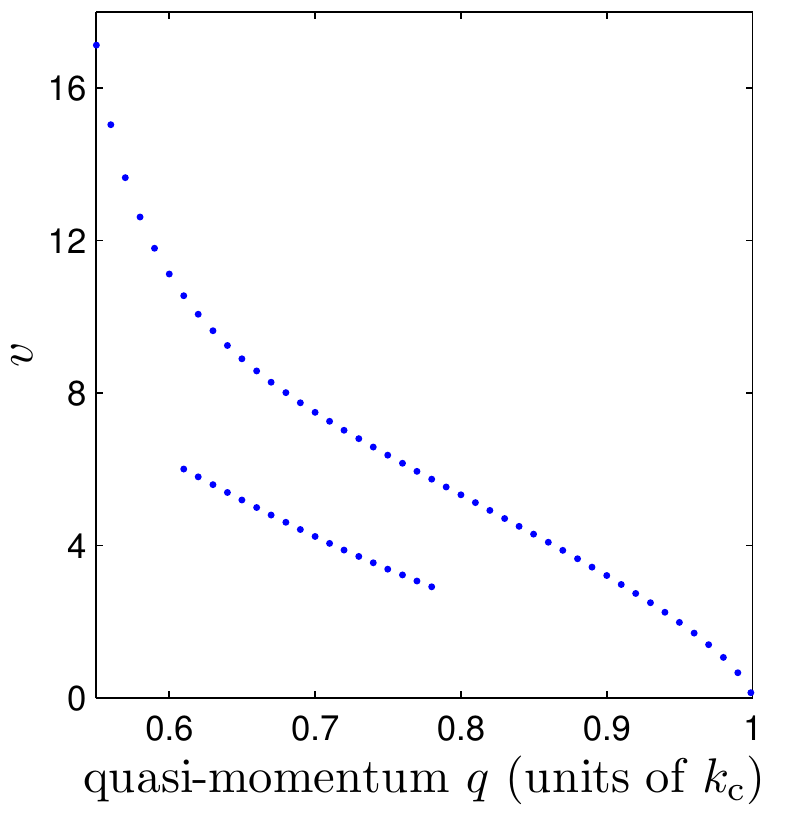} 
\caption{(Color online) A plot showing the values of the state variable $v=U_{0} n_{\mathrm{ph}}$ for which the first three derivatives of the state function $\mathcal{G}$ vanish simultaneously (swallowtail points). For approximately $0.6<q<0.8$  there are two such points for any given $q$.}
\label{fig:orgcenter}
\end{figure}
%%%%%%%%%%%%%%%%%%%%%%%%%%%%%%%%%%%%%%

Taking the derivatives of Eq.\ (\ref{eq:statefnre}), and simplifying slightly the resulting equations, we find 
\begin{align}
 &\left( \frac{\Delta_c}{NU_0} - f \right)^2 -2 v f f^{\rmnum{1}} \left( \frac{\Delta_c}{NU_0} - f\right) =\frac{-1}{N^2U_0^2}
\label{eq:firstder}\\
&\frac{2ff^{\rmnum{1}}+v \left(ff^{\rmnum{2}} + (f^{\rmnum{1}})^2 \right) }{2f^{\rmnum{1}}+vf^{\rmnum{2}}} =
\frac {\Delta_c}{NU_0}\label{eq:secder}\\
&\frac{3ff^{\rmnum{2}} + 3(f^{\rmnum{1}})^2 + v \left( 3f^{\rmnum{1}}f^{\rmnum{2}} +ff^{\rmnum{3}} \right)}{vf^{\rmnum{3}} + 3f^{\rmnum{2}}}
= \frac{\Delta_c}{NU_0}\label{eq:thder}\\
&\frac{12 f^{\rmnum{1}} f^{\rmnum{2}} + 4f f^{\rmnum{3}} + v \left( 4f^{\rmnum{1}}f^{\rmnum{3}} + f f^{\rmnum{4}}  + 3 (f^{\rmnum{2}})^2    
\right)}{vf^{\rmnum{4}} + 4 f^{\rmnum{3}}} = \frac{\Delta_c}{NU_0} \label{eq:fourthder}
\end{align}
where the first, second, third, and fourth derivatives of the function $f(v,q)$ with respect to $v$ are denoted by $f^{\rmnum{1}}$,$f^{\rmnum{2}}$,$f^{\rmnum{3}}$,$f^{\rmnum{4}}$, respectively. In fact, we have  seen Eq.\ (\ref{eq:firstder}) before, as it is the same as Eq.\ (\ref{eq:bistabcondn}) that we used as a condition for bistability. Our strategy will be to find solutions to Eqns (\ref{eq:firstder}--\ref{eq:fourthder}) numerically. In order to facilitate this, observe that \eqnref{eq:secder} and \eqnref{eq:thder} can be combined into a single equation:
\begin{align}
 &\frac{2ff^{\rmnum{1}}+v \left(ff^{\rmnum{2}} + (f^{\rmnum{1}})^2 \right) }{2f^{\rmnum{1}}+vf^{\rmnum{2}}} \nonumber\\ - &\frac{3ff^{\rmnum{2}} +
3(f^{\rmnum{1}})^2 + v \left( 3f^{\rmnum{1}}f^{\rmnum{2}} +ff^{\rmnum{3}} \right)}{vf^{\rmnum{3}} + 3f^{\rmnum{2}}} = 0 \ . \label{eq:thfourthzer}
\end{align}
We solve this equation for $v$ at different values of $q$ by numerically finding the zeros of the left hand side. Once we find the zeros for a particular $q$, we can use \eqnref{eq:secder} to calculate $\Delta_c/NU_0$ at these values. Next, these values of $v$ and $\Delta_c/NU_0$ are used in \eqnref{eq:firstder} to calculate $1/(NU_0)^{2}$. In this last step we find that we obtain the unphysical result $1/(NU_0)^{2}<0$ unless $q>q_{\mathrm{sw}}$, where $q_{\mathrm{sw}} = 0.545$ is a certain critical value of the quasi-momentum. This is illustrated in Fig.\ \ref{fig:u0zerocross} where we plot the values of $1/(NU_0)^2$ computed using the above method for different values of the quasi-momentum $q$ in the neighborhood of $q_{\mathrm{sw}}$. Only the crosses satisfy $1/(NU_0)^{2}>0$. We drop the solutions corresponding to the dots, for which $1/(NU_0)^{2}<0$.
For $q > q_{\mathrm{sw}}$ there are always values of q at which $1/(NU_{0})^2>0$. Note that $q_{\mathrm{sw}} = 0.545$ is a universal result since it does not depend on any other parameter values.

The final step is to check if \eqnref{eq:fourthder} for the fifth derivative  is satisfied at the values of $\Delta_c$, $NU_0$, and $v$ that we computed using the lower derivatives.  We did not find any value
 of $q>q_{\mathrm{sw}}$ where this was the case. The important part of our numerical computation involves the calculation of the derivative of the function $f(v,q)$ for which we do not have an analytical expression. We used the $\mathrm{MATLAB} \textsuperscript{\textregistered}$ \cite{matlab} routine DERIVSUITE \cite{derivsuite} to calculate the derivatives. This routine also provides errors on the derivatives and we can compound errors and find values for expressions like the left hand side of
\eqnref{eq:thfourthzer} with error. We use this error as the tolerance in our zero finding. Some more details of this procedure are provided in the Appendix.

The fact that we did not find a point where the four higher derivatives of the function $\mathcal{G}$ simultaneously vanish  in the range $0<q<1$ (we do not consider negative $q$ because $f(v,q)$  is symmetric under $q \rightarrow -q$) means that although the underlying catastrophe that organizes the solutions is at least a butterfly (because we find five solutions), the four experimental parameters at our disposal $\{\Delta_{c},\eta,q,U_{0}\}$ do not translate into four linearly independent coordinates in control space $\{C_{1},C_{2},C_{3},C_{4}\}$ \cite{note2}. We are therefore not able to navigate freely through the four-dimensional control space and locate the butterfly point at $C_{1}=C_{2}=C_{3}=C_{4}=0$. This is the extension into four dimensions of the situation we already found in Section \ref{sec:catastrophetheoryshallowlattice} for shallow lattices.

\begin{figure}
\includegraphics[width=\columnwidth]{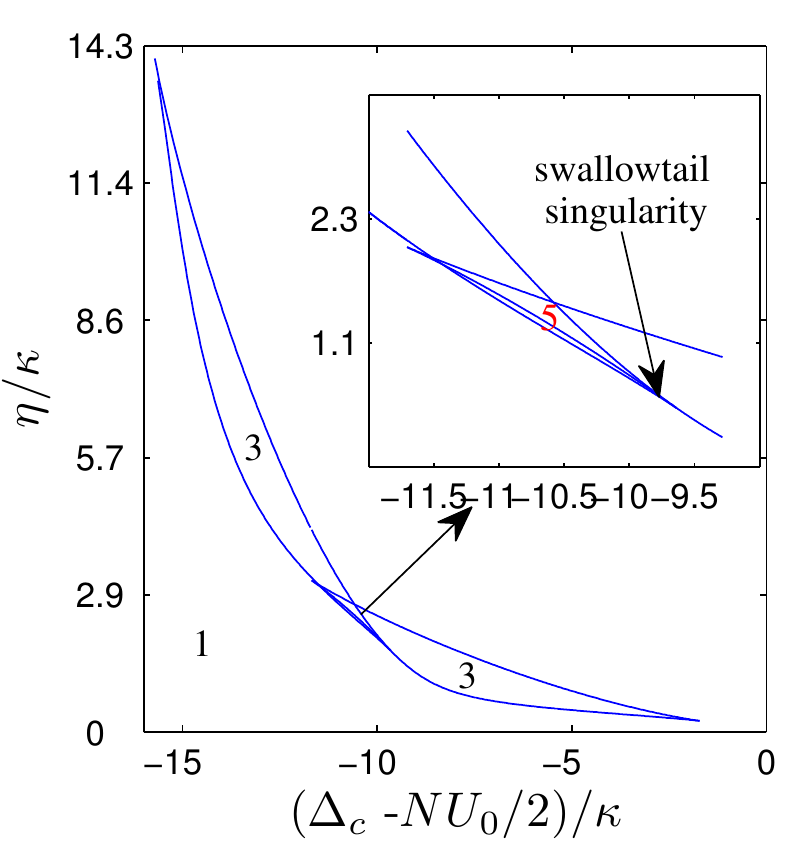} 
\caption{(Color online) Bifurcation structure of the solutions to the self-consistency equation \eqnref{eq:selfconsphnum} in the $\{\eta,\Delta_{c}\}$ plane with $q=0.96$, $U_0 = 1.13 \, \omega_{\mathrm{R}}$, $\kappa = 350 \, \omega_{\mathrm{R}}, N=10^4$, and hence $N U_{0}/2=16.1 \kappa$.  The numbers on the plot indicate the number of solutions for $n_{\mathrm{ph}}$ in that region of the parameter space. The inset shows a swallowtail singularity point where five solutions coalesce into a single solution. The coordinates of the swallowtail point shown in this figure are $v = 0.04$, $\eta = 1.7 \kappa$, $\Delta_{c} = 6.4 \kappa$. }
\label{fig:swallowtail}
\end{figure}

The identification of places where three derivatives of $\mathcal{G}$ simultaneously
vanish means that the highest singularities we have in the parameter space $\{\Delta_{c},\eta,q,U_{0}\}$ are swallowtail points (swallowtails are contained within a greater butterfly catastrophe). In the Appendix we outline a proof that in the neighbourhood of a point where three derivatives of the state equation vanish the potential function must be equivalent to that of a swallowtail catastrophe. Note that these swallowtails occur entirely in control space and are thus true swallow tail catastrophes in the sense of Table \ref{tab:catastrophe}. The swallowtails shown previously in Figures \ref{fig:nonlorenterg}, \ref{fig:diffloops}, \ref{fig:diffloopsU0neg},  \ref{fig:diffloopseta}, \ref{fig:3dplot1} and \ref{fig:3dplot2} are not swallowtail catastrophes because these figures show a combination of state and control spaces. By contrast, Figures \ref{fig:etadelbist}, \ref{fig:etadelmultistab} and \ref{fig:swallowtail} are solely in control space. In particular, Fig.\ \ref{fig:swallowtail} shows a two-dimensional slice through the three-dimensional swallowtail catastrophe in control space. Unlike Fig.\ \ref{fig:etadelmultistab}, this slice includes the swallowtail point, which is the highest singularity on the swallowtail catastrophe, and is the point where four solutions of the state equation (\ref{eq:statefnre}) simultaneously coalesce so that number of solutions changes by $4$ (i.e.\ the point where three derivatives of $\mathcal{G}$ simultaneously vanish). We emphasize that this can only occur when $q>q_{\mathrm{sw}}$. We also note from \figref{fig:orgcenter} that between $0.6<q<0.8$ we find more than one swallowtail point for a given $q$, which is again an indication of the presence of a higher underlying catastrophe.  As described in reference \cite{woodcock}, swallowtails contain two cusps, and butterflies contain two swallowtails.

%%%%%%%%%%%%%%%%%%%%%%%%%%%%%%%%%%%%%%%%

%%%%%%%%%%%%%%%%%%%%%%%%

\section{Stability Analysis}\label{sec10}

The stability of cold atoms in an optical cavity has been treated in Refs.~\cite{nagy09,horak01}. In this case we follow an approach more in line with \cite{machholm03}, where the energy functional, \eqnref{eq:energyfunctional}, and the nonlinear equation of motion, \eqnref{eq:nonlineareqn}, for the atomic wave function serve as the starting points for an examination of energetic, and dynamic stability,
respectively. Hence, we examine the stability of the Bloch states at different values of quasi-momentum at fixed values of $\eta$ and $\Delta_{c}$. Before going into the details of the calculation we note that it is well known from the study of bistability in classical nonlinear cavity systems, that the back-bent branch of the lineshape profile shown in Fig.\ \ref{fig:nonlorent} is unstable, see, for example, \cite{meystre}. From the three dimensional plots Fig.~\ref{fig:3dplot1} and Fig.~\ref{fig:3dplot2}, we see that this back-bent branch corresponds to the upper branch of the loop in energy-quasi-momentum space. Thus, we expect to find this branch unstable. 

We first consider energetic stability in the spirit of \cite{machholm03}. The grand canonical potential per unit length is $G[\psi] = E[\psi] - \mu N$
\begin{align}
 \frac{G[\psi]}{N} &= \frac{1}{\pi }\int _0^{\pi }\text{dx} |\frac{\text{d$\psi
$}}{\text{dx}}|^{^{^2}} \nonumber \\
-&\frac{ \eta ^2}{\text{$\kappa $N}} \arctan \left(\frac{\left(\Delta
_{\mathrm{c}} -\frac{NU_0}{\pi }\int _0^{\pi }\text{dx} |\psi |^{^2}\cos ^2(x)
\right)}{\kappa
}\right)  \nonumber \\ & - \mu \int{dx |\psi(x)|^2}. 
\label{eq:grandcanpotl}
\end{align}
We perturb the wavefunction as $\psi(x) = \psi_0(x) + \delta \psi(x)$, where $\psi_0$ extremizes G, i.e. one of the solutions that we obtained in
Sec.~\ref{sec5}. Since $\psi_0$ is an extremum, the first order variation in $G$ vanishes and the second order contribution can be written as
\begin{align}
\frac{\delta G_2}{N} =  & \langle \delta\psi \vert H_0 \vert \delta\psi \rangle +
\rho  \langle \delta\psi \vert \cos^2(z) \vert \psi_0 \rangle^2
\nonumber \\
 & + \rho \langle \psi_0 \vert \cos^2(z) \vert \delta\psi \rangle^2  \nonumber \\ & + 2\rho
\langle \delta\psi \vert \cos^2(z) \vert \psi_0 \rangle \langle \psi_0 \vert
\cos^2(z) \vert \delta\psi \rangle, \label{eq:secvarG}
\end{align}
where 
\begin{align}
H_0 = -\frac{d^2}{dz^2} + U_0 n_{\mathrm{ph}} \cos^2(z) 
\end{align}
and
\begin{align}
\rho &= \frac{\eta^2 N^2 U_0^2}{\kappa^3}
\frac{\frac{\Delta_{c}-NU_0\langle \psi_0 \vert \cos^2(z) \vert \psi_0
\rangle}{\kappa}}{1+\left( \frac{\Delta_{c}-NU_0\langle \psi_0 \vert
\cos^2(z) \vert \psi_0 \rangle}{\kappa} \right)^2}.
\end{align}
Equation (\ref{eq:secvarG}) can be cast into a simple matrix form~\cite{machholm03}
\begin{align}
\frac{\delta G_2}{N} = \frac{1}{2} \int dx \Psi^{\dagger}(x) A \Psi(x).
\label{eq:ergstab}
\end{align}
Here
\begin{align*}
\Psi(x) = \begin{pmatrix}
            \delta \psi(x) \\ \delta \psi^{*}(x)
           \end{pmatrix}
\end{align*}
and
\begin{align*}
A = \left(\begin{smallmatrix}
      H_0 + 2 \rho \cos^2(x) \psi_0(x) I^{*}[\ldots]  & 2 \rho \cos^2(x)
\psi_0(x) I[\dots]\\
2 \rho \cos^2(x) \psi_0^{*}(x) I^{*}[\dots] & H_0 + 2 \rho \cos^2(x)
\psi_0^{*}(x) I[\ldots]
     \end{smallmatrix} \right),
\end{align*}
where $I[\dots]$ is an integral operator defined by
\begin{align}
I[\delta \psi^{\ast}(x)] \equiv \int \mathrm{d}x \cos^2(x) \psi_0(x) \delta \psi^{\ast}(x).
\label{eq:modecouple}
\end{align}
The eigenvalues of the matrix $A$ decide the energetic stability. If $A$ is positive definite, the solution $\psi_0$ is energetically stable.
%%%%%%%%%%%%
%%%%%%%%%%%%
\begin{figure}
\includegraphics[width=\columnwidth]{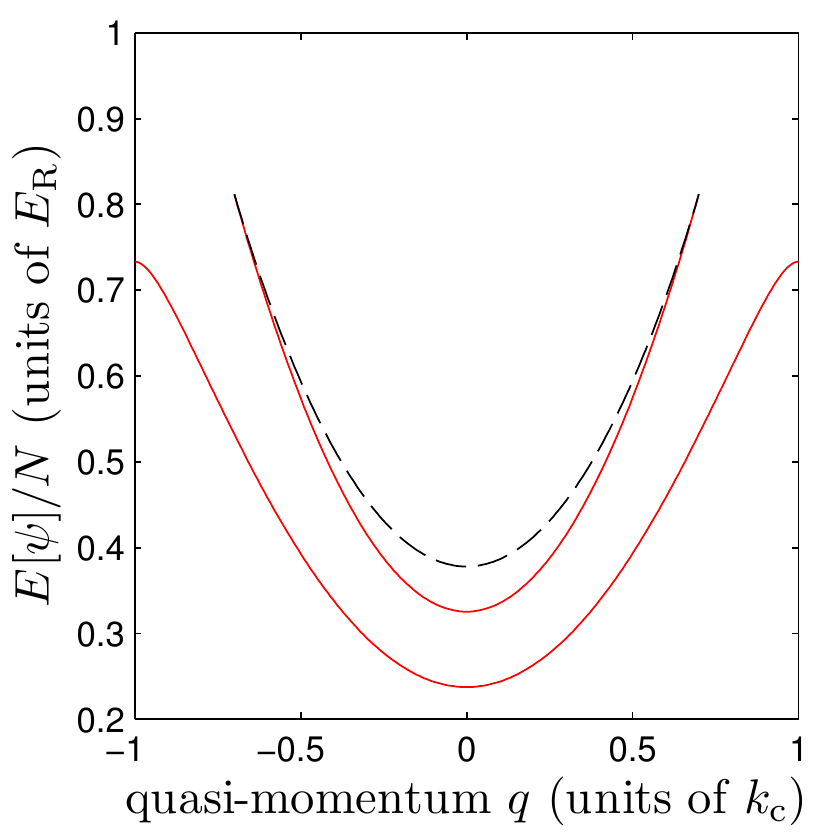}
\caption{(Color online) Energetic and dynamical stability of the band structure loops. The upper branch of the loop (black dashed line) is energetically and
dynamically unstable. The other branches (red
lines) are energetically and
dynamically stable. Parameters are, $\Delta_{c} = 3140 \,
\omega_{\mathrm{R}}$, $\kappa = 350 \, \omega_{\mathrm{R}}$, $U_0 =
\omega_{\mathrm{R}}$, $N=10^4$, $\eta = 2.8 \, \eta_{0}$, and
$\eta_{0} = 325 \, \omega_{\mathrm{R}}$.} 
\label{fig:stabplot}
\end{figure}
%%%%%%%%%%%

In order to examine dynamic stability, we linearize the equation of motion \eqnref{eq:nonlineareqn} by writing $\psi(x,t) = [\psi_0(x) + \delta
\psi(x,t)] e^{-i\mu t}$. This leads to
\begin{align}
&i \frac{\mathrm{d} \delta \psi}{\mathrm{d} t} = \left [ -\frac{\mathrm{d}^2}{\mathrm{d}x^2} + U_0 \mid \alpha_0\mid^2
 \cos^2(x) - \mu \right ] \delta \psi(x) \label{eq:linearisedSE}\\
&+ 2\rho  \cos^2(x) \psi_0(x) \left ( I^{*}[ \delta \psi(x)] + I[
\delta \psi^{*}(x)] \right). \nonumber
\end{align}
One can write a similar equation for $\delta \psi^{*}$ and combine the two into a
matrix equation similar to \eqnref{eq:ergstab}
\begin{align}
i \frac{\mathrm{d} \delta \Psi}{\mathrm{d}t} = \sigma_z A \delta \Psi,
\label{eq:dynstab}
\end{align} 
where $\sigma_z$ is the Pauli $z$-matrix. The solution $\psi_0$ is dynamically stable if all the eigenvalues of $\sigma_z A$ are real. Thus, the
occurrence of complex eigenvalues of $\sigma_zA$ signals dynamical instability. Before we quote the results, a comment is in order about the form
of the perturbations $\delta \psi$. The integral operator in \eqnref{eq:modecouple} couples the perturbation and $\psi_0$. If $\psi_0 =
e^{iqx}U_q(x)$, with $U_q(x)=U_q(x+\pi)$, i.e.\ a Bloch function with quasi-momentum $q$, the form of $\delta \psi$ that leads to non-zero coupling is
\begin{align}
\delta \psi (x) = e^{iqz} \displaystyle\sum_{j} b_j
e^{i2jx}. \label{eq:pertexpansion}
\end{align}
That is, the perturbation should be a Bloch wave with the same quasi-momentum as $\psi_{0}$.
In \eqnref{eq:secvarG} we consider the change in the grand canonical potential per unit length, but the above choice is made to satisfy the requirement that the integral operator $I$ over the system size gives a non-zero answer. A physical way to motivate the above choice goes as follows: in the absence of interatomic interactions the allowed excitations of the quasi-momentum state $q$ have to be in multiples of the crystal momentum $2
k_{c}$ (simply 2 in dimensionless units) since the only source of perturbation is the  interaction with the cavity field. The above form respects this requirement. Also, the number of terms in the Fourier expansion \eqnref{eq:pertexpansion} has to be less than the number of
terms in the original expansion for $\psi_0$ in \eqnref{eq:blochansatz} to avoid spurious instabilities~\cite{machholm03}. Using \eqnref{eq:pertexpansion}, we find that the upper branch of the looped dispersions is both dynamically and energetically unstable as expected. The other two branches are stable. This is shown in \figref{fig:stabplot} for one particular case. We shall not perform the stability analysis for the case when there are five solutions, but anticipate by an extension of the case for the bistability scenario, that two of the solutions will be dynamically unstable and three will be stable.

\section{Summary and Conclusions}\label{sec11}

In this paper we have analyzed bistability in atom-cavity systems in situations where the atoms are in spatially extended states (Bloch waves) with non-zero quasi-momentum $q$. We find that bistability in the number of photons in the cavity goes hand-in-hand with the emergence of loops in the band structure.  Both are manifestations of a bifurcation in the stationary solutions of the coupled atom-light equations of motion.

We have studied how the loops appear and disappear as the laser detuning and the laser pumping rate are changed. In particular, Eq.\ (\ref{eq:etacranalytic}) provides an analytical estimate of the critical pump strength $\eta_{\mathrm{cr}}(q)$ at which bistability sets in. It depends on the quasi-momentum of the atomic state, and predicts that loops first appear at the edges of the first Brillouin zone ($q=\pm 1$) and then move inwards. This is indeed what we find upon solving the coupled atom-light equations numerically: swallowtail loops appear at the edges of the first Brillouin zone as the pump strength $\eta$ is increased above $\eta_{\mathrm{cr}}(q=1)$. As $\eta$ is increased further the swallowtails extend inwards, merge, and detach from the rest of the band to form a separate loop centred at $q=0$ which ultimately closes up and vanishes. A rather similar behaviour is observed as the pump-cavity detuning $\Delta_{c}$ is swept from below the cavity resonance to above it.

The loops we find are qualitatively different from those that occur for BECs in static optical lattices in the presence of repulsive interatomic interactions \cite{wu02,machholm03,mueller02}. There, the loops are centered at the edge of Brillouin zone and cause the dispersion to have a finite slope at that point. By contrast, the band structure we find always has zero slope at the edge of the Brillouin zone. Nevertheless, there are also many similarities, including the stability of the various branches of the loops. We find that the upper branch of the loops are energetically and dynamically unstable,
as expected from optical bistability considerations. 

The extra degree of freedom afforded by the quasi-momentum (over considering only $q=0$) results in  the possibility of tristability, namely regions of parameter space where there are five solutions, three stable and two unstable. The complexity of the solutions in parameter space led us to perform an analysis of the problem in terms of catastrophe theory which is a useful mathematical tool for understanding the organization of bifurcations of solutions. The key to our treatment was the recognition that, because exact solutions for the atomic wave functions are Mathieu functions which are specified only by the lattice depth (once one chooses a quasi-momentum), the photon number $n_{\mathrm{ph}}$ which determines the cavity depth provides a completely equivalent description to the wave function.

In the case of shallow lattices we were able to proceed analytically and found that the structure of the solutions in parameter space when $q=0$ corresponds to a cusp catastrophe, at the most singular point of which three solutions (two stable and one unstable) form a pitchfork bifurcation, and this describes the onset of bistability as the laser pumping is increased. Interestingly, the three experimental parameters $\{\Delta_{c},\eta,U_{0}\}$ reduced to just two effective control parameters.
In the general case of arbitrary lattice depth and $0 \leq q \leq 1$, the highest singularities  we found were swallowtail catastrophes where four solutions simultaneously merge. The swallowtails only exist when $q>q_{\mathrm{sw}}=0.545$. However, there is good evidence that there is an underlying butterfly catastrophe, but, once again, the experimental parameters $\{\Delta_{c},\eta,q,U_{0}\}$ reduced to three effective control parameters meaning that generically one is unable to locate the butterfly point (where five solutions simultaneously merge).

The band structure loops found here have important implications for Bloch oscillations of atoms in cavities \cite{peden09,bpv09}. Bloch oscillations are essentially an adiabatic effect where, as a force is applied to the atoms they remain in the same band but their quasi-momentum evolves linearly in time, as shown in Eq.\ (\ref{eq:BO}). Swallowtail loops in the band structure will have a deleterious effect on Bloch oscillations because, as the quasi-momentum evolves, the atoms will reach the edge of a loop where the branch they are following vanishes. This will lead to a sudden, non-adiabatic, disruption in the state of the atoms as they are forced to jump to another branch or even another band. For BECs in ordinary static optical lattices these non-adiabatic jumps are thought to be the cause of the destruction of superfluidity  during Bloch oscillations \cite{burger01,cataliotti03}. We have not included interactions in our treatment (interactions are necessary for superfluidity), but related effects  will likely occur, especially considering the added heating due to the fact that the lattice depth will also abruptly change at the same point. However, when the loop detaches from the main band it will no longer affect Bloch oscillations.  Furthermore, loops only occur for certain limited regions of parameter space, i.e. inside the cusp and swallowtail catastrophes shown in Figs \ref{fig:etadelbist}, \ref{fig:etadelmultistab}, and \ref{fig:swallowtail}. For experiments involving Bloch oscillations we therefore recommend that parameter regimes are chosen which lie outside to these regions.

Finally, we add that although we have only considered Bloch waves in this paper, localized states (for example Wannier functions) can be formed from superpositions of Bloch waves with different values of the quasi-momentum. In this sense, localized states therefore contain all values of the quasi-momentum and so might be expected to display tristability too. However, it should be borne in mind that the nonlinearity of the system means that superpositions of Bloch states of different $q$ but the same lattice depth will not in general obey the effective Schr\"{o}dinger equation (\ref{eq:nonlineareqn}).

\begin{acknowledgements}
We gratefully acknowledge E.\ A.\ Hinds, D.\ Pelinovsky and M.\ Trupke for discussions. For funding, DHJO and BPV thank the Natural Sciences and Engineering Research Council of Canada, and the Ontario Ministry of Research and Innovation, and JL thanks VR/Vetenskapsr\aa det. 
\end{acknowledgements}

\appendix
\section*{Appendix}\label{app}

We shall now sketch a proof showing that the function $\mathcal{G}$ defined in \eqnref{eq:statefnre} produces swallowtail catastrophes between $0.545 \leq q \leq 1$  (in fact it produces two lines of swallowtail points, as shown in \figref{fig:orgcenter}) where its derivatives up to third order vanish simultaneously  \cite{poston,gaite98}. 

Consider, for example, the point in control and state space given by  $C_0 =
\{ \Delta_c = 0.90,  \eta=14.5, q = 0.69, U_0 = 0.15 \}$ and $ v_0=7.75, n_\mathrm{ph} = 50.3$ (frequencies are measured in units of $\kappa$ and the number of atoms is set at $N=10^2$). 
The numerical package \cite{derivsuite} we used for calculating the derivatives gives error bounds allowing us to estimate the accuracy of our calculations. For the point $\{C_{0}, v_{0} \}$ we found that the right hand side of Eq.\ (\ref{eq:thfourthzer}) was equal to $8.09 \times 10^{-15}$ with an error of $4.00 \times 10^{-13}$. This means that the third derivative of the state function vanishes within error, indicating a swallowtail point. However, the smallest value we found anywhere in parameter space for the difference between the left hand side and the right hand side of Eq.\ (\ref{eq:fourthder}) was $-8.64 \times 10^{-5}$ with error $6.00 \times 10^{-12}$. This means that the fourth derivative of the state function does not vanish within error, suggesting there is no butterfly point. The value of the quasi-momentum at the point where we found this minimal fourth derivative was  $q = q_{\mathrm{sw}} = 0.545$.

Before outlining the proof, we first give some definitions \cite{poston}. If $n$ is the number of state variables, then consider a function $p$: $\mathcal{R}^n \rightarrow \mathcal{R}$
\begin{itemize}
\item $j^k p$ is the Taylor expansion of $p$ to order $k$
\item $J^k p$ is $j^k p$ minus its constant term
\item $p$ is $k$-deteminate at $0$ if any smooth function $p+q$, where $q$ is of order $k+1$ (leading order term of the Taylor expansion is of order $k+1$), can be locally expressed as $p(y(s))$ with $y$: $\mathcal{R}^n \rightarrow \mathcal{R}^n$ being a smooth reversible change of co-ordinates.
\item $E_n^k$ is the vector space of polynomials in $s_1,\dots,s_n$ of degree $\leq k$. 
\item $J_n^k$ is the subspace of $E_n^{k}$ with zero constant term
\item $\Delta_k(p)$ is the subspace of $J_n^k$ spanned by all $ \overline{Q j^{k}\left (\frac{\partial p}{\partial s_i} \right)}^k$, where $1\leq i \leq n$, $Q \in E_n^k$, and the bar symbol represents the restriction to the $k$-th order of the expansion.
\item The codimension of a function $p$ is the codimension of $\Delta_k(p)$ in $J_n^k$ for any $k$ for which $p$ is $k$-determinate.  
\item An $r-\mathrm{unfolding}$ of $p$ at $0$ is a function:
\begin{align*}
& P:\mathcal{R}^{n+r} \rightarrow \mathcal{R},\\
 & \left( s_1,\dots,s_n,t_1,\dots,t_r\right)  \mapsto P(s,t) = P_t(s),
\end{align*}
such that $P_{0,...0}(s) = p(s)$.
\end{itemize}
More informally, the term ``unfolding'' refers to how the catastrophe unfolds as one moves away from the origin in control space. At the origin in control space the catastrophe reduces to its most singular part, known as its \emph{germ}. For example, from Table \ref{tab:catastrophe}, the germ of the swallowtail catastrophe is given by $s^5$. The terms in the potential function which depend on the control parameters are called the unfolding terms, and the number of them is equal to the codimension. If $P$ is an $r-\mathrm{unfolding}$ of $p$, set
\begin{align}
 & \left \{ w_1^k(P),\dots,w_r^k(P) \right \} =  \nonumber \\ 
& \left \{ \frac{\partial}{\partial t_1}  \left( J^k (P_{t_1,0,\dots,0})\right),\dots, \frac{\partial}{\partial t_r}
\left( J^k (P_{0,\dots,t_r})\right)            \right \}.
\end{align}
$W^k(P)$ is the subspace of $J_n^{k}$ spanned by $\{w_1^k(P),\dots,w_r^k(P) \}$.

Referring to the standard forms given in Table \ref{tab:catastrophe},  the potential function and state equation for the swallowtail are given by
\begin{align}
 \Phi(s;C) &= \frac{1}{5}s^5+\frac{C_{3}}{3} s^3 + \frac{C_{2}}{2} s^2 + C_{1}s \label{eq:thomswpot}\\
\mathcal{G} (s;C) &\equiv s^4 + C_{3} s^2 + C_{2} s+ C_{1} = 0  \ . \label{eq:thomswstate}
\end{align}
Notice that the state equation for a swallowtail catastrophe is similar to the potential function for a cusp
catastrophe up to a constant $C_{1}$. Since the state function $\mathcal{G}$ is the central object in the treatment given in Section \ref{sec9} rather than the potential function, instead of proving that the underlying potential function is equivalent to that of a swallowtail catastrophe, we
will prove that the state function $\mathcal{G}$ around the singular point $v_0$ and $C_0$ is equivalent to the potential function of a cusp catastrophe (note that this is different from the small photon number case we studied in Section \ref{sec:catastrophetheoryshallowlattice} where we showed that the underlying potential was equivalent to the potential for a cusp catastrophe). To that end, first notice that the role of the constant term $C_{1}$ in \eqnref{eq:thomswstate} is played by $-\eta^2 U_0$ in \eqnref{eq:statefnre}. Subtracting this function we have a modified form of $\mathcal{G}$ (where we have also dropped the dependence on $q$ since we are focusing on a particular quasi-momentum):
\begin{align}
F(v;\{\Delta_c, U_0 \}) = F(v;C) = v  + v(\Delta_c-NU_0 f(v,q))^2
\label{eq:modG}
\end{align}
which satisfies $F^{'}(v_0;C_0) = F^{''}(v_0;C_0) = F^{'''}(v_0;C_0) = 0$. In order to have a function defined in the neighborhood of $v_0$ and $C_0$, let us set the origin of $v$ at $v_0$ and the origin of $C$ at $C_{0}$ and define
\begin{align}
F_1(v;\{ \Delta_c, U_0\}) \equiv F(v+v_0,C+C_0) - F(v_0,C_0).
\label{eq:F1fn}
\end{align}
Thus, we have $F_1(0,0) = 0$ and for the function $g(v) \equiv F_1(v,\{0,0,0\})$ the most singular point is at $v=0$ where
$g^{'},g^{''},g^{'''}$ vanish. The function $g$ is  the germ which we described above, and is the key feature which identifies the catastrophe. When  $g$ is Taylor expanded around $0$ one has
\begin{align}
g(v) = \frac{g^{(\rmnum{4} )}}{4!} v^{4} + \frac{g^{(\rmnum{5} )}}{5!} v^{5}+ \mathcal{O}(v^{6})
\label{eq:gfn}
\end{align}
where $g^{(\rmnum{4} )}(0)$ is the first non-zero Taylor coefficient. This means that $g$ is $4$-determined around $0$ and we say that $g \sim v^4$ around $0$. According to Table \ref{tab:catastrophe}, the canonical unfolding of the $4$-determined germ around $0$ is the cusp catastrophe $\Phi(s;C) = s^4/4+ C_{2}s^2/2 + C_{1} s  $, where $v$ and $s$ are related via a differomorphism (smooth transformation of coordinates). 

Next we calculate the codimension of $g$. The Jacobian ideal for $g$ is in this case $\Delta_4(g) = \{ v^4,v^3 + \frac{g^{\rmnum{5}}}{4g^{\rmnum{4}}}\vert_{v = 0} v^4 \}$. Hence, the codimension of $g$ is $\mathrm{dim}(J_1^4)-\mathrm{dim}(\Delta_4(g)) = 4-2 = 2$.
The function $F_1$ is thus a $2$ parameter unfolding of the germ $g$. In order to prove that the function $F_1$ can be described by a cusp catastrophe, we need to prove that $F_1$ is isomorphic as an unfolding to the canonical form $\Phi(s;C) = s^4/4+ C_{2}s^2/2 + C_{1} s $. In order to do this we need to invoke the idea of transversality. 

Transversality generalizes what we know of two intersecting lines in a two dimensional plane to multidimensional manifolds. Two subspaces of a manifold are transverse if they meet in a subspace that is as small in dimension as possible. If $X_1$ (dim $r$) and
$X_2$ (dim $t$) are subspaces of $X$ (dim $n$), $X_1$ and $X_2$ are transverse if their intersection is empty or if it is of the dimension max$(0,r+t-n)$. Our first aim is to prove that the $2-\mathrm{unfolding}$ of $F_1$ is a versal unfolding. To do this we use a defining theorem for versality from \cite{poston} which states that: an $r-\mathrm{unfolding}$ $P$ of $p$, where $p$ is $k$-determinate is versal if and only if  $W^{k}(P)$ and $\Delta_k(p)$ (defined above) are
transverse subspaces of $J_n^k$. We have already found $\Delta_4(g)$, the polynomial space $W^4(F_1)$ is spanned by the vectors:
\begin{align*}
w_1(F_1) &=  \frac{\partial}{\partial U_0} \left( J^4(F_1(0,{U_0,0,})) \right),\\
w_2(F_1) & = \frac{\partial}{\partial \Delta_c} \left( J^4(F_1(0,{0,\Delta_c}))\right),\\
\end{align*}
The expressions depend on the derivatives of the coupling function $f(v,q)$ and the value of the parameters at the singular point $\{v_0,C_0 \}$. They are too cumbersome to state here but their general forms are given by
\begin{align*}
 w_i (F_1) = \displaystyle\sum_{j=1..4}{z_{ij}v^{j}},
\end{align*}
which we determined numerically and all of the $z_{ij}$'s are non-zero. The polynomials $w_i$ are linearly independent which gives the dimensionality  $\mathrm{dim} (W^4(F_1)) = 2$.  Furthermore, we have verified that the rank of the matrix formed by the polynomial coefficieints  of $\Delta_4(g)$ and $W^4(F_1)$ is $4$  and this combined with the fact that $\mathrm{dim}(\Delta_4(g))+\mathrm{dim}(W^4(F_1)) = 2 + 2 = \mathrm{dim}(J_1^4)$ proves that $\Delta_4(g)$ and $W^4(F_1)$ are
transverse. Thus, by the theorem stated above $F_1$ is a versal unfolding of the germ $g$ and since it is a $2-\mathrm{unfolding}$ (codimension of $g=2$) it is also universal \cite{poston}. This proves the equivalence of the unfolding of $F_1$ to the cusp catastrophe.


\begin{thebibliography}{15}
\bibitem{note2}{Another possibility is that the butterfly point is exactly located at the point $1/(NU_0)^2 = 0$, and $q$ can take any value. This is a rather unphysical situation because, assuming a finite number of atoms, it can only occur when $U_{0} \rightarrow \infty$ and this then means $v \rightarrow \infty$. From Fig. \ref{fig:cos2xfn}, we see that when $v \rightarrow \infty $ the function $f(v,q)$ becomes very flat and so all its derivatives vanish allowing Eqns (\ref{eq:firstder})-(\ref{eq:fourthder}) to be satisfied in a rather trivial way. Therefore, we ignore this solution.}
\end{thebibliography}
\end{document}